\documentclass[fleqn,a4paper,authoryear]{elsarticle}

\usepackage[a4paper,margin=1in]{geometry}
\usepackage{fancyhdr}
\usepackage{dsfont}
\usepackage[utf8]{inputenc}
\usepackage[fleqn]{amsmath}
\usepackage{amssymb}
\usepackage{booktabs}
\usepackage{longtable}
\usepackage{xcolor}
\usepackage{appendix}
\usepackage{stmaryrd}
\usepackage{subfig}
\DeclareMathOperator{\erfc}{erfc}
\usepackage{mathtools}
\usepackage{amssymb}
\usepackage{xcolor}
\usepackage{graphicx}
\graphicspath{ {/home/alexandrosstathas/Documents/PhD_topics/PAPER_III/Sumission_Papaer_III_Arxiv/} }
\usepackage{float}
\usepackage{layout} 
\usepackage{appendix}

\usepackage{hyperref}
\usepackage{xr-hyper}
\title{Fault friction under thermal pressurization during large coseismic-slip Part II: Expansion to the model of frictional slip}

\begin{document}

\author[1]{Alexandros Stathas}
\author[1]{Ioannis Stefanou\corref{cor1}}
\ead{ioannis.stefanou@ec-nantes.fr}
\cortext[cor1]{Corresponding author}
\address[1]{Institut de Recherche en Génie Civil et Mécanique (UMR CNRS 6183), Ecole Centrale de Nantes, Nantes, France}

\begin{abstract}
\noindent
In \cite{Alexstathas2022a} we presented the frictional response of a bounded fault gouge under large coseismic slip. We did so by taking into account the evolution of the Principal Slip Zone (PSZ) thickness using a Cosserat micromorphic continuum model for the description of the fault's mechanical response. The numerical results obtained differ significantly from those predicted by the established model of thermal pressurization during slip on a mathematical plane (see \cite{Mase1987,Rice2006,Platt2014} among others). These differences prompt us to reconsider the basic assumptions of a stationary strain localization on an unbounded domain present in the original model. We depart from these assumptions, extending the model to incorporate different strain localization modes, temperature and pore fluid pressure boundary conditions. The resulting coupled linear thermo-hydraulic problem, leads to a Volterra integral equation for the determination of fault friction. We solve the Volterra integral equation by application of a spectral collocation method (see \cite{tang2008spectral}), using Gauss-Chebyshev quadrature for the integral evaluation. The obtained solution allows us to gain significant understanding of the detailed numerical results of Part I. We investigate the influence of a traveling strain localization inside the fault gouge considering isothermal, drained boundary conditions for the bounded and unbounded domain respectively. We compare our results to the ones available in \cite{lachenbruch1980frictional,Lee1987,Mase1987} and \cite{Rice2006}. Our results establish that when a stationary strain localization profile is applied on a bounded domain, the boundary conditions lead to a steady state, where total strength regain is achieved. In the case of a traveling instability such a steady state is not possible and the fault only regains part of its frictional strength, depending on the seismic slip velocity and the traveling velocity of the shear band. In this case frictional oscillations increasing the frequency content of the earthquake are also developed. Our results indicate a reappraisal of the role of thermal pressurization as a frictional weakening mechanism.
\end{abstract}
\maketitle

\section{Introduction \label{PartII}}
\begin{figure}[h!]
  \centering
  \includegraphics[width=0.45\linewidth]{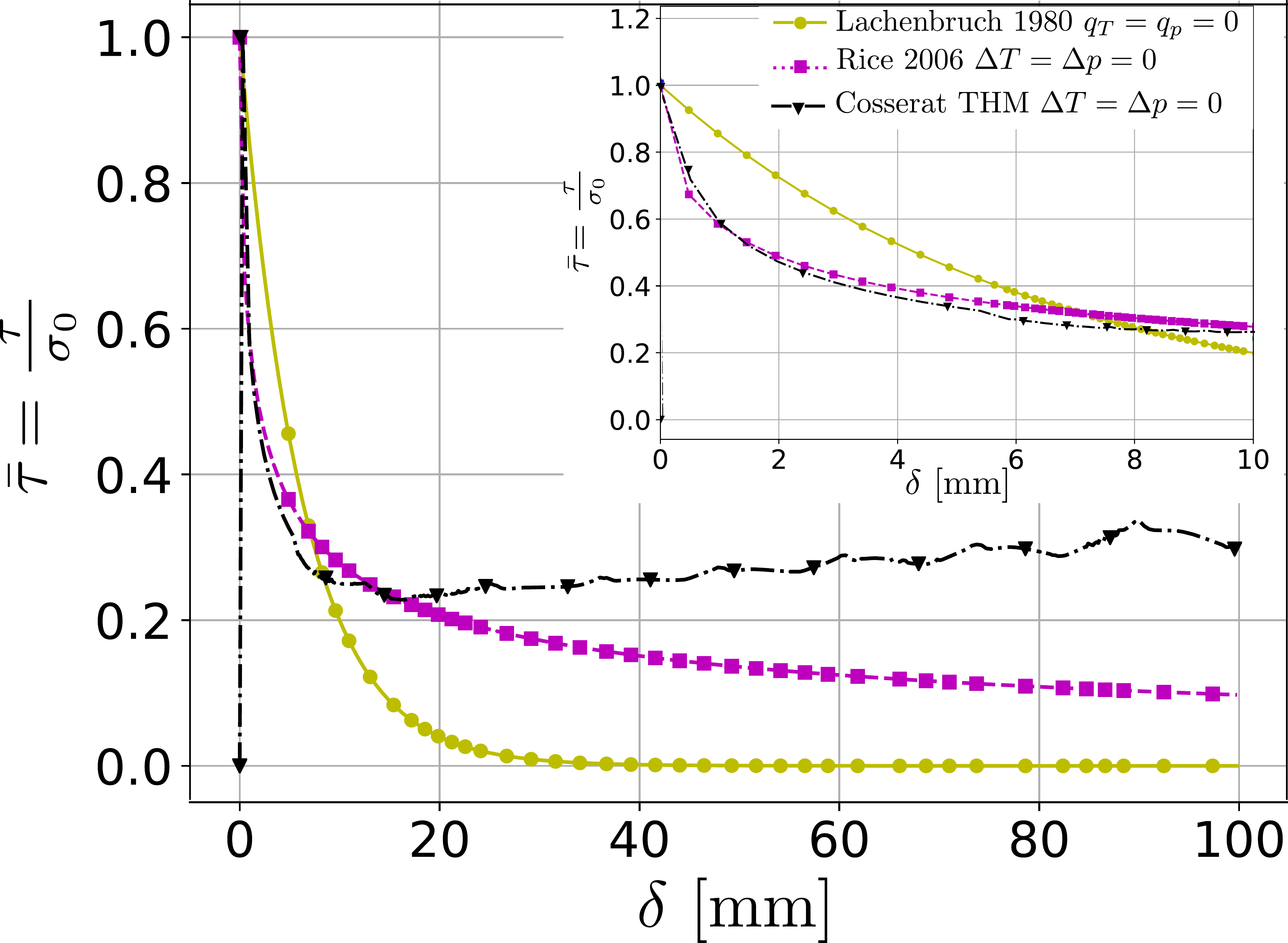}
  \caption{Comparative normalized friction $\bar{\tau}$- displacement $\delta$ results. The purple-square curve presents the frictional response of the established thermal pressurization model in the case of slip on a mathematical plane under isothermal drained boundary conditions lying at infinity \protect{\cite{Mase1987,Rice2006}}. The yellow-circle curve presents the frictional response of the established model when uniform slip occurs under adiabatic undrained boundary conditions for a fault gouge of height H=1 mm under shear velocity V=1 m/s \cite{lachenbruch1980frictional}. The black-triangle line corresponds to the frictional response of the micromorphic model of Part I, \cite{Alexstathas2022a} for the same fault gouge under isothermal drained boundary conditions. For small values of slip $\delta\leq 10$ mm, the response of the micromorphic model lies between the two limiting cases, however, it diverges as seismic slip $\delta$ accumulates.}
  \label{ch: 6 fig: classical_vs_micromorphic_compare}  
\end{figure}

\noindent The results of Part I \citep[see][]{Alexstathas2022a}, concerning the influence of the weakening mechanism of thermal pressurization, diverge -spectacularly- from the expected behavior based on the model of \cite{Mase1987,Rice2006}. Furthermore, we note that these results, indicate the divergence to take place long before the completion of the seismic slip $\delta$. This holds true for the range of commonly observed seismic slip velocities $\dot{\delta}\;\in\;\{0.1\sim 1\}$ m/s and seismic slip displacements ${\delta}\;\in\;\{0.1\sim 1\}$ m (see \cite{Harbord2021,Rempe2020}). In this follow-up paper, Part II, we investigate the reasons for this divergence between the theoretical results and their implications for the appreciation of thermal pressurization as one of the main weakening mechanism during coseismic slip. Our investigation leads us to extend the existing model of slip on a mathematical plane by relaxing its key assumptions.\\
\newline
\noindent In Figure \ref{ch: 6 fig: classical_vs_micromorphic_compare} we compare the frictional response of the micromorphic model used in Part I \citep[see][]{Alexstathas2022a}, with the response of the established model for the limiting cases of uniform shear \cite{lachenbruch1980frictional} and shear on a mathematical plane \cite{Mase1984,Rice2006}. In particular, the two limiting responses of the established model, depend on the width of accumulating strain localization inside the fault gouge, which we call the Principal Slip Zone (PSZ). They are characterized respectively by: a) the uniform slip across the fault gouge (see \cite{lachenbruch1980frictional}), and b) the localization of slip on a mathematical plane (see \cite{Lee1987,Mase1987,rempel2006effects,Rice2006}). We note that while at the initial stages of slip (see inset of Figure \ref{ch: 6 fig: classical_vs_micromorphic_compare}) the response of the micromorphic model lies inside the the envelope of the limit cases, at larger values of slip it diverges, presenting frictional regain and the initiation of frictional oscillations. These results come in contrast to the strictly monotonous behavior predicted by the limiting cases of uniform slip and slip on a mathematical plane.  
\newline
\noindent We note here, that the limiting cases are predicted by the established model of thermal pressurization, under three important assumptions (see \cite{lachenbruch1980frictional,Mase1987,Rice2006}): First of all, the thickness of the yielding region, which corresponds to the PSZ, coincides with the fault gouge. Prescribing the thickness, and therefore, the shape of plastic strain profile, essentially decouples the mechanical and thermo-hyraulic components of the coupled THM problem (see \cite{Mase1984}). Secondly, the variability between the thermal and hydraulic parameters of the gouge and the surrounding rock is assumed to be small, thus the thermo-hydraulic boundaries for the THM coupled problem lie at infinity. In essence the change of hydrothermal parameters between the fault gouge and the surrounding rock is neglected. Lastly, the position of the heat source due to the dissipation inside the PSZ, remains stationary inside the fault, and coincides to the position of the fault gouge.\\
\newline
\noindent These assumptions, however, are not representative of observations. We know from laboratory experiments and in situ observations that the fault gouge, has a finite thickness of the order of some milimeters, and it does not deform in a uniform manner (see \cite{myers2004evolution,Brantut2008}). In fact, inside the fault gouge, the principal slip zone (PSZ) is a region of finite thickness of the order of some micrometers depending on the geomaterial’s microstructure (see \cite{Muhihaus1988,sibson1977fault}). In this configuration the fault gouge and the region that accumulates the majority of the plastic deformation inside it -the PSZ- do not coincide. Furthermore, one needs to acknowledge, that the frictional response inside the fault gouge is dependent on the ratio of thermal to hydraulic diffusivity of the fault gouge and the surrounding rock. In particular we know from the works of \cite{aydin2000fractures,passelegue2014influence,yao2016crucial} that the hydraulic and thermal diffusivity of the gouge is smaller than that of the surrounding rock by 1 to 2 orders of magnitude. This large difference between the parameters of the fault system needs to be accounted for. Finally, there is experimental evidence of fault gouges, that are thicker than expected according to the existing models of \cite{Platt2014} and \cite{Rice2006}, and of closely adjacent fault gouges, see \cite{nicchio2018development}, whose existence can be linked to the possibility that the position of the PSZ is not stationary, rather it is traveling inside the fault gouge, possibly expanding the latter in the process. \\
\newline
\noindent There is also theoretical evidence considering the possibility of a traveling PSZ inside the fault gouge. In this case the preferred mode of strain localization might not be that of the divergence kind described in \cite{Rice1975}, rather it can be a ``flutter'' type instability, corresponding to a traveling strain localization profile (PSZ). According to the Lyapunov theory of stability (see \cite{lyapunov1893problem,Brauer1969}), a traveling strain localization (PSZ) is manifested by the appearance of a Lyapunov exponent with imaginary parts. The transition form a stationary instability of divergence type to a flutter traveling instability is called a Hopf bifurcation. For more details we refer to section \ref{Part I-ch: 5 Traveling_Instabilities} of Part I, \cite{Alexstathas2022a}, where we have shown numerically that, for stress states common in faults, traveling instabilities are present in the linear stability analysis for Cosserat continua under strain softening and apparent softening due to multiphysical couplings. In the broader context of a classical continuum, under hydraulic couplings, \cite{Benallal2003} have shown that traveling instabilities are also present. It is not yet clear if this is the case for a classical continuum under THM couplings (see \cite{benallal2005localization}).\\
\newline
\noindent In this paper we provide an explanation concerning the differences between the numerical results of Part I (see \cite{Alexstathas2022a}) and the frictional response predicted by limit cases of the classical model \cite{lachenbruch1980frictional,Mase1987,Rice2006}. To this end we expand the classical model of thermal pressurization described in \cite{Rice2006}, and extend its applicability to cases of bounded fault gouges and traveling strain localization modes of the PSZ. We will use the same thermal, hydraulic and geometric parameters for the fault gouge as in the model of Part I, \cite{Alexstathas2022a}. Next, we will collapse the PSZ, where yielding and frictional heating takes place, onto a mathematical plane by employing the same formalism used in \cite{Lee1987,Mase1987,Rice2006}. We assume further, that the yield (dissipation) obeys a Coulomb friction law with the Terzaghi normal effective stress. The mechanical behavior of the layer outside the yielding plane is ignored and for the purposes of this model it can be considered as rigid. This allows us to avoid solving a BVP for the mechanical part, which significantly simplifies the problem (c.f. \cite{Alexstathas2022a}).\\
\newline  
\noindent The decision to collapse the PSZ onto a mathematical plane can be justified based on the results of Part I, see sections \ref{Part I-ch: 5 sec: Two_step_procedure}, \ref{Part I-ch: 5 viscosity_reference} and Figures \ref{Part I-ch: 5 fig: tau_u_velocity_compare-fit},\ref{Part I-ch: 5 fig: R_bstar_compare}). These lead us to the observation that it is the hydraulic and thermal parameters of the fault that mainly affect thermal pressurization. We note, however, that the Cosserat radius, which is a parameter connected with the grain size and the material properties of the granular medium is still an indispensable internal length for the numerical analyses of Part I because: a) it assures the mesh independence of the numerical results, and b) it provides finite localization width over which frictional heating takes place. However, for the analyses performed in this part (Part II), the introduction of the Dirac delta distribution prescribing the profile of the plastic strain rate, thus decoupling the mechanical and thermohydraulic component of the propblem of thermal pressurization, allows us to overcome the problem of incorporating the microstructure to the model, considerably simplifying the analysis. This allows us to elaborate on the effect of the boundary conditions on the frictional response. Furthermore, this simplification allows us to gain further insight into the problem, because the mechanisms responsible for the principal characteristics of the response of the micromorphic model described in Part I (restrengthening, frictional oscillations) can be isolated and investigated separately, corroborating the numerical results of Part I.\\
\newline
\noindent This paper is structured as follows: In section \ref{ch: 6 sec: section 2} we present the basic equations of the classical model of thermal pressurization (see Mase1987,Mase1984,Rice2006) and our proposed expansion to the cases of bounded fault gouges and a traveling PSZ, by elaborating further on their differences. Our extended model leads to a Volterra integral equation of the second kind, which cannot be solved analytically as in the case of \cite{Rice2006}. 
 In section \ref{ch: 6 sec: section 3}, we solve the Volterra integral equation of the second kind by applying a Spectral Collocation Method with Lagrange basis functions (SCML), based on the work of \cite{Evans1981,Elnagar1996,tang2008spectral}. This is a general spectral method and can handle the challenging task of integrating the Volterra equation under different assumption of boundary conditions and traveling strain localization modes, when other analytical approaches such as Laplace transform, Adomian decomposition Method and Taylor series expansion fail \cite[see][]{wazwaz2011linear,boyd20062000}.\\
\newline
\noindent Having described our model and the solution procedure, we present in section \ref{ch: 6 sec: section 4} a series of applications showcasing the differences with the analyses in \cite{Rice2006}. The applications include the frictional responses of: (a) a stationary PSZ on a bounded isothermal drained domain, (b) a moving PSZ on an unbounded isothermal drained domain, and (c) a moving PSZ on a bounded isothermal drained domain. The original solution in \cite{Rice2006} is obtained as a special case of the more general solutions presented here and is taken as reference (see Figure \ref{ch: 6 fig: frictional_behavior_stationary_unbounded}).\\
\newline 
\noindent Finally, in conclusions we discuss the implications of our results concerning the introduction of a traveling PSZ inside the fault gouge. Our results are important as they describe better the underlying physical process of seismic slip. Moreover, a traveling PSZ naturally enhances the frictional response with oscillations, which in turn can enhance the ground acceleration spectra with higher frequencies as observed in nature \cite{Aki1967a,BRUNEJN1970}. Moreover, our results are valuable in the context of experiments for the description of the weakening behavior due to thermal pressurization (see \cite{Badt2020}), for controlling the transition from steady to unsteady slip and for the nucleation of an earthquake (see \cite{Rice1973a,viesca2015ubiquitous}). They are also central in earthquake control, as they provide bounds for the apparent friction coefficient with slip and slip-velocity enabling modern control strategies (see \cite{Stefanou2019,Stefanou2020,tzortzopoulos2021Thesis}.
\section{Thermal pressurization model of slip on a plane \label{ch: 6 sec: section 2}}
\subsection{Problem statement}
\begin{figure}[h!]
  \centering
  \includegraphics[width=0.45\linewidth]{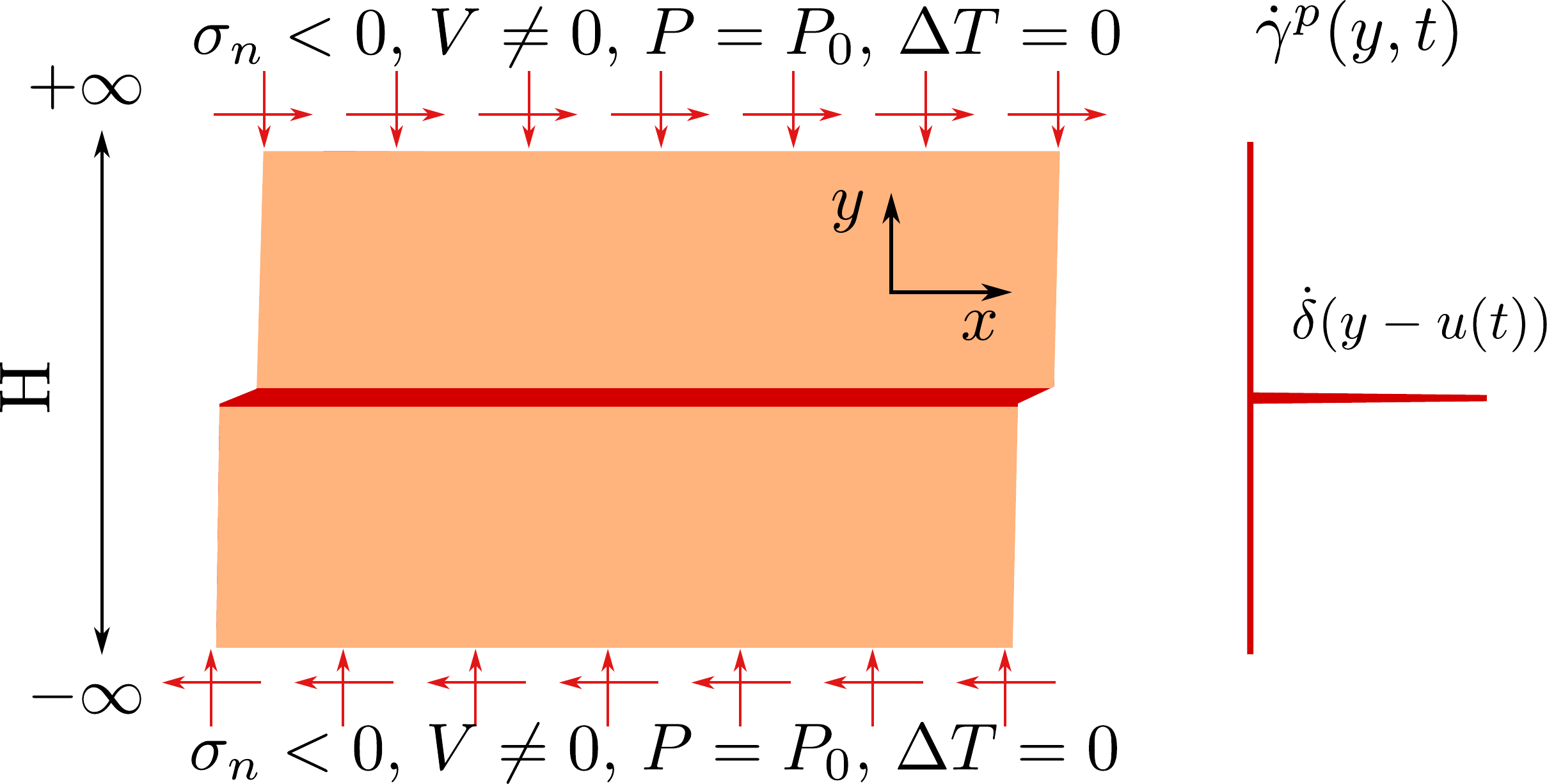}
  \caption{The established model of thermal pressurization: The values of Pressure and temperature are prescribed at infinity. The bodies outside the fault gouge (red color) are considered rigid. Deformation loclalizes on a mathematical plane and the PSZ coincides with the fault gouge.}
   \label{ch: 6 fig: classical_model_fault_gouge}  
\end{figure}

\noindent We already discussed in the introduction, that the current model of thermal pressurization, shown in Figure \ref{ch: 6 fig: classical_model_fault_gouge}, assumes that yielding is constrained on a mathematical plane inside the domain, which is modeled based on the Coulomb friction criterion (see equation \eqref{ch: 6 eq: Coulomb friction} below). This plane will be also called yielding plane in the following. Contrary to \cite{Mase1987,Rice2006}), the yielding plane is  not considered stationary inside the domain. Instead its position $u(t)$ is allowed to change with a velocity $\dot{u}(t)=v(t)$. Furthermore, we will not consider only isothermal drained boundary conditions lying at infinity. In particular, we will also take into account the case, where the fault gouge is bounded under isothermal drained boundary conditions lying at $y=0\;y=\text{H}$. 
\\
\noindent At the yielding region inside the layer heat is produced due to dissipation, $D$. The thermal source then is described by the calculation of the plastic work rate, $\dot{D}$, at the position of the failure plane, namely:
\begin{align} 
\dot{D}=\tau(t)\dot{\gamma}^p(y,t).
\end{align} 
\noindent In the above formula, friction $\tau(t)$, which is the main unknown of the problem, is independent of position $y$, due to equilibrium considerations along the height of the fault gouge ($\frac{\partial \tau}{\partial y}=0$, in the absence of inertia, see also \cite{Rice2006}). The term $\dot{\gamma}^p(y,t)$ is the plastic strain rate inside the fault gouge. In the established model of thermal pressurization this term is prescribed with the help of a Dirac distribution stationed at the plane of symmetry, $y=0$ (see \cite{Mase1984,Rice2006}). Here we expand the term $\dot{\gamma}^p(y,t)$ to account for a traveling PSZ at position $y=u(t)$ as follows: 
\begin{align}
\dot{\gamma}^p(y,t)=V(t)\delta_{\text{Dirac}}(y-u(t)). \label{ch: 6 plastic_strain_rate_prf}
\end{align}
\noindent In the case of $u(t)=0$, no traveling can take place and the stationary condition of \cite{Mase1987,Rice2006} is recovered. In the model of \cite{Rice2006} the author considers that the shear rate $V(t)$ applied at the boundaries of the fault gouge is constant $V(t)=V$. We adopt this assumption although seismic slip rate during coseismic slip may vary significantly (see \cite{Rempe2020}).
The equations of the established model \cite{Mase1987,Rice2006} are then written as follows:
\begin{align}
&\tau(t)=f(\sigma_n-P_{max}(t)), \text{ on the yielding plane,} \label{ch: 6 eq: Coulomb friction}\\
&\frac{\partial \Delta T}{\partial t}=c_{th}\frac{\partial \Delta T}{\partial y^2}+\frac{1}{\rho C}\tau(t)V\delta_{\text{Dirac}}(y-u(t)),\\
&\frac{\partial \Delta P}{\partial t}=c_{hy}\frac{\partial \Delta P}{\partial y^2}+\Lambda\frac{\partial \Delta T}{\partial t},\\
&\Delta T\|_{y=0,H}=\Delta P\|_{y=0,H}=0,\\
&\Delta T(y,0)=\Delta P(y,0)=0,\;P(y,t)=\Delta P(y,t)+P_0,
\end{align}
\noindent where $f$ is the friction coefficient, $c_{th},\;c_{hy}$ are the thermal and hydraulic diffusivity of the layer (same values for the fault gouge and the fault walls) respectively, $\rho C$ is the specific heat density of the layer, $V$ is the shearing rate of the layer, assumed here to be constant, and $\Lambda=\frac{\lambda^\star}{\beta^\star}$ is the thermal pressurization coefficient (see Table \ref{ch: 6 table:material_properties}). The symbol $(\|_\alpha)$ indicates the value of temperature and pressure fields at position $\alpha$ of the model, while $P_0$ is the ambient value of pore fluid pressure at the boundaries of the fault gouge. We note that if we set the boundary conditions at infinity (i.e. $\|_{\pm\infty}$) the boundary assumptions of \cite{Mase1987} and \cite{Rice2006} are recovered.\\
\newline
\noindent We note here, that prescribing the position of the yielding plane $y=u(t)$ implies that the position of $P_{max}$ is known, and coincides with the position of the thermal load. Thus the above model is valid if the position of the maximum pressure $P_{max}(t)$ and the yielding plane coincide. In this case, because the yielding position is prescribed and the plastic strain profile known, the mechanical behavior is decoupled and the resulting coupled thermo-hydraulic problem described above is linear.\\
\newline 
\noindent Applying the pore fluid pressure solution (see also equation \eqref{ch: 6 eq: pressure_sol_mod} in \ref{Appendix A}) to the failure criterion results finally to the following Volterra integral equation of the second kind for the determination of the layer’s frictional response under constant shearing rate (see \cite{Rice2006,wazwaz2011linear}):
\begin{align}
&\tau(t)=f(\sigma_n-P_0)-\frac{f\Lambda V}{\rho C(c_{hy}-c_{th})}\int_{0}^t\tau(t^\prime)G^\star(y,t;y^\prime,t^\prime,c_{hy},c_{th})\|_{y=y^\prime}dt^\prime,\label{Volterra_integral_equation_mod}
\end{align} 
\newline
\noindent where $G^\star(y,t;,y^\prime,t^\prime,c_{hy},c_{th})$ is the kernel of the integral equation , which we present further in section \ref{ch: 6 sec: BD_loading_presentation} (see also  \cite{cole2010heat}). The kernel indicates the influence of the thermal load applied at position $y’$ and time $t’$ in the pore fluid pressure observed at position $y$ and time $t$. Throughout our analysis we make the assumption that the maximum value of pore fluid pressure $P_{max}(t)$, at observation time $(t)$, lies at the point of application of the thermal load $y^\prime$. This assumption is then verified numerically. Hence, the position of observation of $P_{max}(t)$, $y$, is equal to $y=y^\prime$, and the kernel $G^\star(y,t;y^\prime,t^\prime,c_{hy},c_{th})$ needs to be calculated at $y=y^\prime$.\\
\newline
\noindent We note that the frictional response is dependent on the strain localization mode and the boundary conditions applied at the fault gouge. The first influences the form of the thermal load as a function of time and position, while the latter influences the form of the kernel of the coupled linear thermo-hydraulic problem at hand. For the purposes of our analyses we will consider the cases of: 1) an unbounded fault gouge under a) a stationary PSZ, described in \cite{Rice2006}, b) a traveling PSZ at a constant velocity $v$, and 2) a bounded fault gouge under a) a stationary PSZ, and b) a traveling PSZ, where position is a periodic function of time i.e. $y^\prime=u(t^\prime )$ (see equation \eqref{ch: 6 plastic_strain_rate_prf} and section \ref{ch: 6 sec: section 4}). The periodic movement of the PSZ is justified on the basis of the numerical analyses presented previously in Part I (see \cite{Alexstathas2022a}, Figures \ref{Part I-ch: 5 fig: tau_u_velocity_compare_1},\ref{Part I-ch: 5 fig: l_dot_gamma_final_1000-T_p_final_1000}.) We present the relevant Green's function kernels in section \ref{ch: 6 sec: BD_loading_presentation}. In order to solve the modified resulting Volterra integral equation \eqref{Volterra_integral_equation_mod}, we have employed the collocation quadrature method described in \cite{tang2008spectral} as explained in section \ref{ch: 6 sec: section 3}.
\\
\newline
\noindent Having defined the differences between the classical and the extended model of thermal pressurization described in this section, we comment further on the differences between our linear extended model and the one used in the fully nonlinear analyses of Part I, \cite{Alexstathas2022a}. In particular, in Part I, a micromorphic model together with THM couplings, was used for the determination of the PSZ thickness during coseismic slip. The application of a micromorphic continuum leads to a finite thickness for the PSZ, which guarantees mesh objectivity of the numerical results. Because the thickness of the PSZ is finite, the thermal load applied inside the PSZ is distributed over the PSZ thickness. Furthermore, the finite thickness of the PSZ is a crucial part of the mechanism explaining the appearance of traveling PSZ inside the fault gouge as we have argued in Part I. We further note that the yield criterion employed in the analyses of Part I was a Drucker-Prager yield criterion, while, here we make use of a Mohr-Coulomb yield criterion. The use of the Mohr Coulomb criterion allows us to describe the friction $\tau(t)$ with the help of the normal stress $\sigma_n$ to the yielding plane, instead of the combination of normal stresses required in the case of the Drucker-Prager.
\subsection{Cases of Interest \label{ch: 6 sec: BD_loading_presentation}}
\noindent 
 We consider four cases for the loading and boundary conditions concerning the evaluation of the fault friction during coseismic slip. We first separate between stationary and traveling modes of strain localization and then we further discriminate between unbounded and bounded domains in order to cover all possible cases. The separation of the fault's frictional response into these categories leads to four different expressions for the Green's function kernel $G^\star(y,t;y^\prime,t^\prime,c_{hy},c_{th})$ in equation \eqref{Volterra_integral_equation_mod}.\\
\newline
\noindent Here we will provide the analytical expressions, for the kernels to be substituted into equations \eqref{Volterra_integral_equation_mod}. In naming the Green's function kernels we used the subscript naming conventions of \cite{cole2010heat}. Namely for diffusion in 1D line segment domains, the letter $X\alpha\beta$ is adopted, where $\alpha,\;\beta$ are the left $(y=0)$ and right $(y=\text{H})$ boundaries of the domain respectively. They can take the values $0$ or $1$ indicating an unbounded or a bounded domain respectively, under homogeneous Dirichlet boundary conditions.\\
\newline
\noindent We begin by introducing the Green's function kernels of the unbounded $X00$ and the bounded $X11$ cases in the case of a 1D diffusion equation under homogeneous Dirichlet boundary conditions.\\
\newline
For the unbounded case we use:
\begin{align}
G_{X00}(y,t;y^\prime,t^\prime,c) = \frac{1}{2\sqrt{\pi c(t-t^\prime)}}\exp{\left[-\frac{(y-y^\prime)^2}{4 c(t-t^\prime)}\right]}. \label{ch: 6 eq: G_X00_kernel}
\end{align}
\noindent Similarly for the bounded case we use:
\begin{align}
G_{X11}(y,t;y^\prime,t^\prime,c) = \frac{2}{L}\sum_{m=1}^{\infty}\exp{\left[-m^2\pi^2c\frac{t-t^\prime}{\text{H}^2}\right]}\sin{\left(m\pi\frac{y}{\text{H}}\right)}\sin{\left(m\pi\frac{y^\prime}{\text{H}}\right)}.\label{ch: 6 Green's_long_co-time}
\end{align}
\noindent We note here that $c$ can be either $c_{th}$ or $c_{hy}$ depending on the diffusion problem in question. The kernels $G^\star_{X\alpha\beta}(y,y^\prime,t-t^\prime,c_{hy},c_{th})$ of the pressure diffusion problem based on the impulse of the frictional response for given boundary strain localization modes and boundary conditions are given by:
\begin{itemize}
\item[•] Stationary mode of strain localization
\begin{itemize}
\item[•]Unbounded domain, $\alpha=0,\;\beta=0,\;x^\prime=0$, \cite[see][]{Rice2006}
\begin{align}
G^\star_{X00}(y,t;0,t^\prime,c_{hy},c_{th})=c_{hy}G_{X00}(y,t;0,t^\prime,c_{hy})-c_{th}G_{X00}(y,t;0,t^\prime,c_{th}).\label{ch:6 Green's_unbounded_stationary}
\end{align}
\item[•]Bounded domain $\alpha=1,\;\beta=1,\;,x^\prime=0$
\begin{align}
G^\star_{X11}(y,t;0,t^\prime,c_{hy},c_{th}) =\ c_{hy}G_{X11}(y,t;,0,t^\prime,c_{hy})-c_{th}G_{X11}(y,t;0,t^\prime,c_{th}).
\end{align}
\end{itemize}
\item[•] Traveling mode of strain localization
\begin{itemize}
\item[•]Unbounded domain, $\alpha=0,\;\beta=0,\;y^\prime=u(t^\prime)$:
\begin{align}
\begin{aligned}
G^\star_{X00}(y,t;y^\prime,t^\prime,c_{hy},c_{th})=c_{hy}G_{X00}(y,t;u(t^\prime),t^\prime,c_{hy})-c_{th}G_{X00}(y,t;u(t^\prime),t^\prime,c_{th}).
\end{aligned}
\end{align}
\item[•]Bounded domain, periodic trajectory in time, $\alpha=1,\;\beta=1,\;y^\prime=u(t^\prime)$:
\begin{align}
\begin{aligned}
G^\star_{X11}(y,t;y^\prime,t^\prime,c_{hy},c_{th}) =\ c_{hy}G_{X11}(y,t;u(t^\prime),t^\prime,c_{hy})-c_{th}G_{X11}(y,t;u(t^\prime),t^\prime,c_{th}).
\end{aligned}\label{ch:6 Green's bounded periodic trajectory}
\end{align}
\end{itemize}
\end{itemize}  
\section{Methods for the numerical solution of linear Volterra integral equations of the second kind \label{ch: 6 sec: section 3}}
\noindent The solution of linear integral equations of the second kind can be sought with a variety of different analytical and numerical methods. From an analytical standpoint, these methods include methods from operational calculus namely, Laplace, Fourier or $\mathcal{Z}$-Transform \cite[see][]{churchill1972operational,brown2009complex,Mavaleix-Marchessoux2020}, the use of Taylor expansions for the integrand inside the integral and the method of Adomian decomposition \cite[see][]{wazwaz2011linear,Evans1981}. The case of a stationary yielding mathematical plane described in \cite{Rice2006} can, and has been solved analytically, making use of the Laplace transform. Those methods depend on the convolution property of the integral in the integral equation to transform it into a simpler algebraic equation. The challenge then lies in the inversion of the relation obtained in the auxiliary (frequency) domain back to the time domain. However, as the complexity of the Green's function kernels and the loading function increases due to the introduction of boundary conditions and different assumptions concerning the trajectory of the shear band along the fault gouge, such an inversion is not always possible analytically. We are then forced to use numerical methods for the solution of the above Volterra integral equation.\\
\newline
\noindent The above analytical methods have also their numerical counterparts, with the use of the Discrete Fourier Transform (DFT) being a central part in most numerical solution procedures. However, use of the DFT is most efficient when the integral equation to be solved has the form of a convolution. This is not always the case in our problem. For instance, the kernel described in equation \eqref{ch:6 Green's bounded periodic trajectory} has terms in $(t,\;t^\prime)$ that do not involve their difference $(t-t^\prime)$, and therefore, its use in equation \eqref{Volterra_integral_equation_mod} results in the integral term not being a convolution. In order to handle the above difficulty then, we will make use of another class of numerical methods called spectral collocation methods, which solve the integral equation \eqref{Volterra_integral_equation_mod} directly in the time domain. These methods are conceptually easy to use, and since no inversion is required, they are able to handle very general cases of Green's function kernels and loading functions.\\
\newline
\noindent In what follows, we will make use of the Spectral Collocation Method with Lagrange basis functions (SCML) for the numerical solution of the integral equation \eqref{Volterra_integral_equation_mod} \cite[see][and section \ref{ch: 6 sec: section 3.2}]{tang2008spectral,Elnagar1996}. The SCML method will be shown to handle both the bounded and unbounded domains and the cases of stationary vs traveling strain localization.  

\subsection{Collocation method\label{ch: 6 sec: section 3.2}}
\noindent We begin by normalizing equation \ref{Volterra_integral_equation_mod}. We choose the following normalization parameters $t_0=\frac{\text{H}^2}{c_{th}},\;\tau_0=f(\sigma_n-p_0),y_0=\text{H},r_c=\frac{c_{hy}}{c_{th}}$. The normalized equation is the given by:
\begin{align}
\bar{\tau}(\bar{t})=1-\frac{f\Lambda V}{\rho C}\frac{ \text{H}}{c_{th}(r_c-1)}\int^{\bar{t}}_0\bar{\tau}(\bar(t)^\prime)\bar{G}^{\star}(\bar{y},\bar{t};\bar{y}^\prime,\bar{t^\prime})\|_{y=y^\prime}d\bar{t^\prime}\label{ch 6: eq: normalized_integral_equation}
\end{align}
\noindent where $\bar{\tau}=\frac{\tau}{\tau_0},\;\bar{t}=\frac{t}{t_0},\bar{t}^\prime=\frac{t^\prime}{t_0}\;\bar{y}=\frac{y}{y_0},\bar{y}^\prime=\frac{y^\prime}{y_0}$ and $\bar{G}^\star(\bar{y},\bar{t};\bar{y}^\prime,\bar{t^\prime})$ is the normalized Green's function kernel given by:
\begin{itemize}
\item[•] In the unbounded case:
\begin{align}
\bar{G}_{X00}^\star(\bar{y},\bar{t};\bar{y}^\prime,\bar{t^\prime})=\frac{1}{2}\left[\frac{r_c}{\sqrt{\pi r_c(\bar{t}-\bar{t}^\prime)}}\exp\left[-\frac{(y-y^\prime)^2}{4r_c(\bar{t}-\bar{t}^\prime)}\right]-\frac{1}{\sqrt{\pi(\bar{t}-\bar{t}^\prime)}}\exp\left[-\frac{(y-y^\prime)^2}{4(\bar{t}-\bar{t}^\prime)}\right]\right]\label{ch 6: eq: normalized_unbounded_kernel}
\end{align}
\item[•] In the bounded case:
\begin{align}
\bar{G}_{X11}^\star(\bar{y},\bar{t};\bar{y}^\prime,\bar{t^\prime})=2\left[r_c\sum_{m=1}^\infty\exp\left[-(m\pi)^2r_c(\bar{t}-\bar{t}^\prime)\right]-\sum_{m=1}^\infty\left[-(m\pi)^2r_c(\bar{t}-\bar{t}^\prime)\right]\right]\sin{\left(m\pi\bar{y}\right)}\sin{\left(m\pi\bar{y}^\prime\right)}\label{ch 6: eq: normalized_bounded_kernel}
\end{align}
\end{itemize}
\noindent Based on the work of \cite{tang2008spectral}, we apply a spectral collocation method for the calculation of the frictional response described by equation \eqref{ch 6: eq: normalized_integral_equation}. Spectral methods allow for evaluation of the solution in the whole domain of the problem yielding exponential degree of convergence (see \cite{tang2008spectral}). The principle of the method is the substitution of the unknown function 
$\bar{\tau}(\bar{t})$ inside the integral equation by a series of polynomials that constitute a polynomial basis. We then opt for the minimization of the residual between the exact and the approximate solution at specific collocation points inside the problem's domain. Here we use the Chebyshev orthogonal polynomials of the first kind (see \cite{trefethen2019approximation}). Because the Chebyshev polynomial of the first kind constitute a basis in the interval [-1,1], we transform the integral equation \eqref{ch 6: eq: normalized_integral_equation} to lie in this interval (see Appendix \ref{Appendix H}).The integral equation then reads:
\begin{align}
U(\bar{z})=1-\frac{f\Lambda V}{\rho C}\frac{ \text{H}}{c_{th}(r_c-1)}\frac{\bar{\text{T}}}{2}\int^{\bar{z}}_{-1}U(s)G^{\star}\left(\bar{y},\frac{\bar{\text{T}}}{2}(\bar{z}+1);\bar{y}^\prime,\frac{\bar{\text{T}}}{2}(s+1)\right)ds,\label{ch 6: eq: normalized_integral_equation_mew_interval}
\end{align}
\noindent where we note that $U(\bar{z})=\bar{\tau}(\frac{\bar{\text{T}}}{2}(\bar{z}+1))$. In the previous equation we performed a change in the integration variable from $\bar{t}\in [0,\frac{\bar{\text{T}}}{2}(\bar{z}+1)]$ to $s\in[-1,\bar{z}]$ so that the unknown function $U(s)$ inside the integral remains in the same form as $U(\bar{z})$ outside the integral.
\noindent Next, we choose to approximate the unknown function in equation \eqref{ch 6: eq: normalized_integral_equation_mew_interval} (i.e. frictional response) by its Lagrange interpolation i.e:
\begin{align}
U(\sigma)\approx \sum_{j=0}^{N}{} U(\bar{z}_j)F_j(\sigma)
\end{align}
\noindent The Lagrange interpolation allows a function to be approximated as a linear combination of the Lagrange cardinal polynomials $F_j(\sigma)$, and weights $U(\bar{z}_j)$ corresponding to the values of the function at specific points $z_{j}$. The Lagrange cardinal polynomials have the property that $F_m(\bar{z}_n)=\delta_{mn}$, where $\delta_{mn}$ is the kronecker symbol. We choose to express the Lagrange polynomials with the help of the Chebyshev polynomials of the first kind, and we choose the set of approximation nodes $\bar{z}_j$ to correspond to the extrema of the Chebyshev polynomial of the first kind, of degree $N$ (see \cite{trefethen2019approximation}). In this case the interpolating polynomial is written as follows:
\begin{align}
&U(\sigma)\approx \sum_{j=0}^{N}{}^\prime U(\bar{z}_j)P_j(\sigma),\\
&P_j(\sigma)=\begin{cases}
\frac{(-1)^{j}}{\sigma-\bar{z}_j}/\sum\limits_{k=0}^{N}{}^\prime\frac{(-1)^k}{\sigma-\bar{z}_k}&\sigma\neq \bar{z}_j\text{ or } \sigma\neq \bar{z}_k\label{ch: 6 eq: barycentric formula}\\
2&\sigma=\bar{z}_0\text{ or }\sigma=\bar{z}_N\\
1&\sigma=\bar{z}_j\text{ and }j\neq 0\text{ or }j\neq N\\
0&\sigma=\bar{z}_k\\
\end{cases}\\
&\sum_{j=0}^{N}{}^\prime(\cdot)_j=\sum_{j=0}^N(\cdot)_j-\frac{(\cdot)_0+(\cdot)_N}{2}\label{ch: 6 eq: mod sum}
\end{align}
\noindent where the barycentric formula involving the modified sum $\sum\limits_{j=0}^{N}{}^\prime(\cdot)$ is used for the cardinal polynomials and the interpolation (see \cite{trefethen2019approximation}). By making use of the barycentric formula in equation \eqref{ch: 6 eq: barycentric formula} we are able to evaluate the cardinal polynomials fast and with smaller error that other conventional approaches (see \cite{trefethen2019approximation,tang2008spectral}). We note that the Lagrange interpolation polynomial at the selection of Chebyshev points $\{\bar{z}_i\}$ stays unaffected by Runge's phenomenon. Runge's phenomenon is the observation that the high polynomial degree Lagrangian interpolation in equidistant grids leads to high error at the approximation of points that don't belong to the set of interpolation nodes. The effect is more pronounced near the boundaries of the interpolation domain. \\
\newline
\noindent For the numerical evaluation of the integral in equation \eqref{ch 6: eq: normalized_integral_equation_mew_interval} the Clenshaw-Curtis quadrature will be used since it is compatible with the iterpolation nodes used. We note here that the choice of the interpolation nodes $\bar{z}_j$ -extrema of the degree N Chebyshev polynomial of first kind- leads to quadrature weights of positive sign, which reduces the error of the summation. If equidistant points were used as a quadrature rule of high order $(N>7)$, this would lead to quadrature weights of alternating sign increasing the integration error \cite{Quarteroni2007}.
 We transform once again the integral of equation \eqref{ch 6: eq: normalized_integral_equation_mew_interval} from $s\in[-1,z]$ to $\theta\in[-1,1]$ in order to apply the appropriate quadrature rule for integration. the new integral equation reads:
\begin{align}
U(\bar{z})=1-\frac{f\Lambda V}{\rho C}\frac{ \text{H}^2}{c_{th}(r_c-1)}\frac{\bar{\text{T}}}{2}\frac{\bar{z}+1}{2}\int^{1}_{-1}U(s(\bar{z},\theta))G^{\star}\left(\bar{y},\frac{\bar{\text{T}}}{2}(\bar{z}+1);\bar{y}^\prime,\frac{\bar{\text{T}}}{2}(s(\bar{z},\theta)+1)\right)ds,\label{ch 6: eq: normalized_integral_equation_mew_interval1}
\end{align} 
\noindent
The discretized form of equation \eqref{ch 6: eq: normalized_integral_equation_mew_interval1} for the Clenshaw Curtis quadrature scheme is given by:
\begin{align}
&U(\bar{z}_i)=1-a\bar{t}_i\sum_{j=0}^{N}{}^\prime U_j(\bar{z}_j)\sum_{p=0}^{N}P_j(s_{ip})G^{\star}\left(\bar{y},\bar{t}_i;\bar{y}^\prime,\bar{t}^\prime_{ip})\right)w_p,\label{ch 6: eq: normalized_integral_equation_mew_interval_discretized_full}
\end{align}
\noindent where,$
s_{ip}=s(\bar{z}_i,\theta_p),\;\bar{t}_i=\frac{\bar{z_i}+1}{2}\bar{\text{T}},\;\bar{t}^\prime_{ip}=\frac{\bar{\text{T}}}{2}\left(s_{ip}+1\right),\;a=\frac{f\Lambda V}{\rho C}\frac{ \text{H}^2}{c_{th}(r_c-1)}\frac{\bar{\text{T}}}{2}$.\\
\newline
\noindent 
Finally, by adopting the indicial notation with summation over repeated indices our system is written as:
\begin{align}
&(\delta_{i,j}+A^\star_{i,j})U_j(\bar{z}_j)=g_i,
\end{align}
\noindent where $g_i=1$ and $A^\star_{i,j}$ is given by:
\begin{align}
&A^\star_{i,j}=A_{ij}-B_{ij}\\
&A_{ij}=a\bar{t}_i \sum_{p=0}^{N}P_j(s_{ip})G^{\star}\left(\bar{y},\bar{t}_i;\bar{y}^\prime,\bar{t}^\prime_{ip})\right)w_p\\
&B_{ij}=\begin{cases} 
A_{i0}&,\;j=0\\
A_{iN}&,\;j=N\\
A_{ij}&,\;j\neq 0 \text{ or } j\neq N\\
\end{cases}
\label{ch: 6 eq: algebraic_system}
\end{align}
\noindent or in matrix form:
\begin{align}
\left(I+A^\star\right)U=G,
\end{align}
\noindent We can then solve the algebraic system to find the interpolation coefficients $U_j$ of the numerical solution. Due to the properties of the Lagrange polynomials the coefficients $U_i$ are also the values of the numerical solution at the specific times $t_i$.
\section{Applications \label{ch: 6 sec: section 4}}
\noindent In this section we will present the evolution of the frictional strength $\tau(t)$ for the different cases of loading and boundary conditions described in section \ref{ch: 6 sec: BD_loading_presentation}. The available values for the fault gouge properties considered homogeneous along its height are given in Table \ref{ch: 6 table:material_properties}.
\begin{table}[h!]
\begin{center}
\begin{tabular}[]{l l l l l l}
\hline
Parameters & Values & Properties & Parameters & Values & Properties \\
\hline
\hline
$f$ &$0.5$ &- &$\Lambda$ & $2.216$ & MPa/$^o$C \\
$\sigma_n$ & $200$ &MPa &$\rho C$ & $2.8$ &MPa$/^o$C \\
$P_0$ & $66.67$ &MPa &$c_{hy}$ & $10 $ &mm$^2$/s \\
$\text{H}$ & $1$ &mm &$c_{th}$ & $1 $ &mm$^2$/s \\
\hline
\end{tabular}
\caption{Material parameters of a mature fault at the seismogenic depth \protect\cite[see][]{rice2006heating,Rattez2018b}.}
\label{ch: 6 table:material_properties}
\end{center}
\end{table}
\subsection{Stationary strain localization mode}
\subsubsection{Stationary strain localization on an unbounded domain \label{ch: 6 sec: Rice_proc_inf_layer}}
\begin{figure}[h!]
  \centering
  \includegraphics[width=0.75\linewidth]{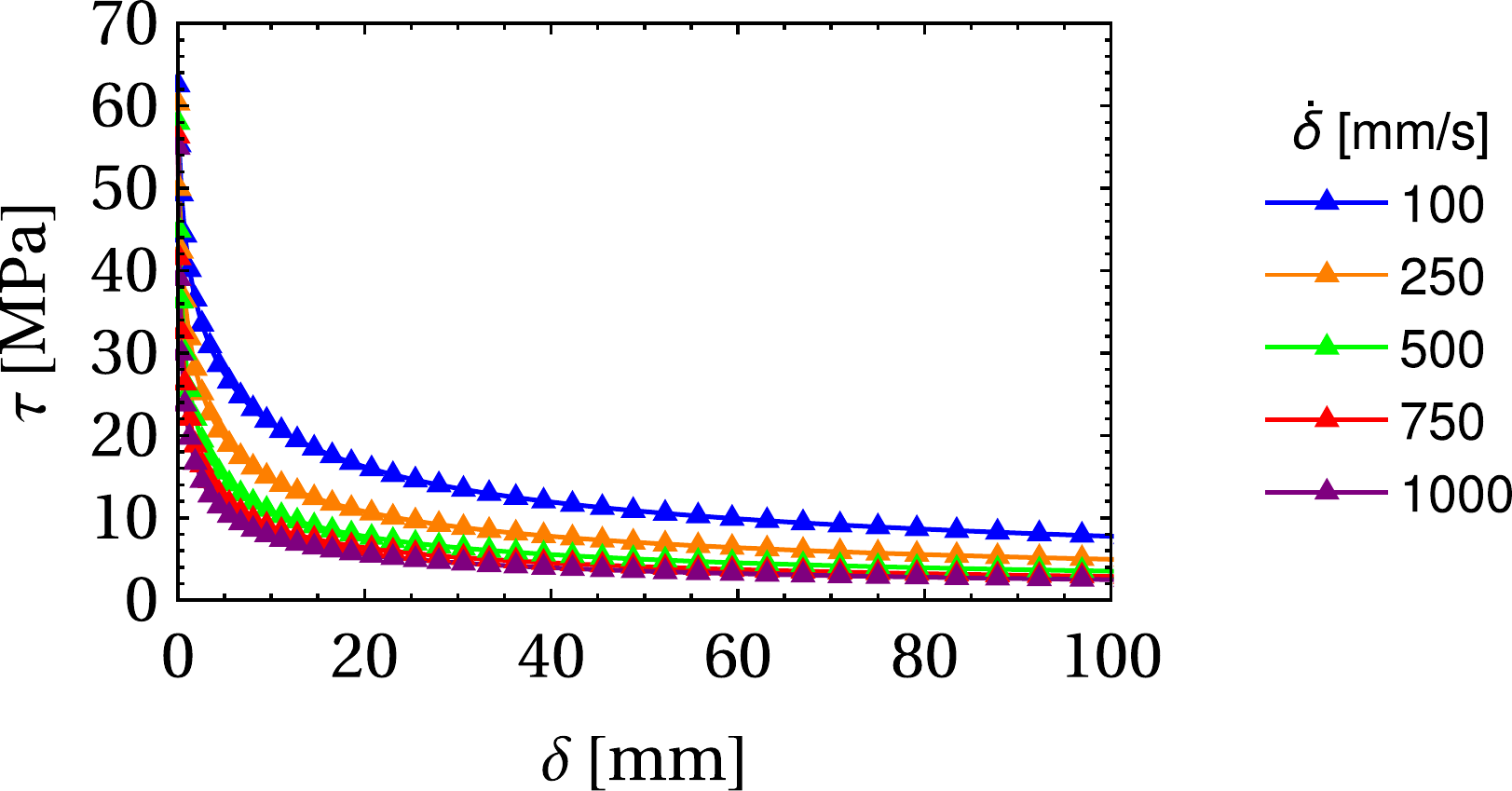}
  \caption{Left: $\tau-\delta$ response of the layer for different slip velocities $\dot{\delta}$ applied. Due to the constant isothermal drained conditions at the boundary near infinity the solution tends asymptotically to the zero steady state solution. For different values of the velocity $\dot{\delta}$, the analytical solution is presented by a continuous line and the numerical solution is presented by the triangle markers. The numerical solution obtained by the SCLM method, coincides to the analytical curves.}
  \label{ch: 6 fig: frictional_behavior_stationary_unbounded}  
\end{figure}

\begin{figure}[h!]
  \centering
  \includegraphics[width=0.75\linewidth]{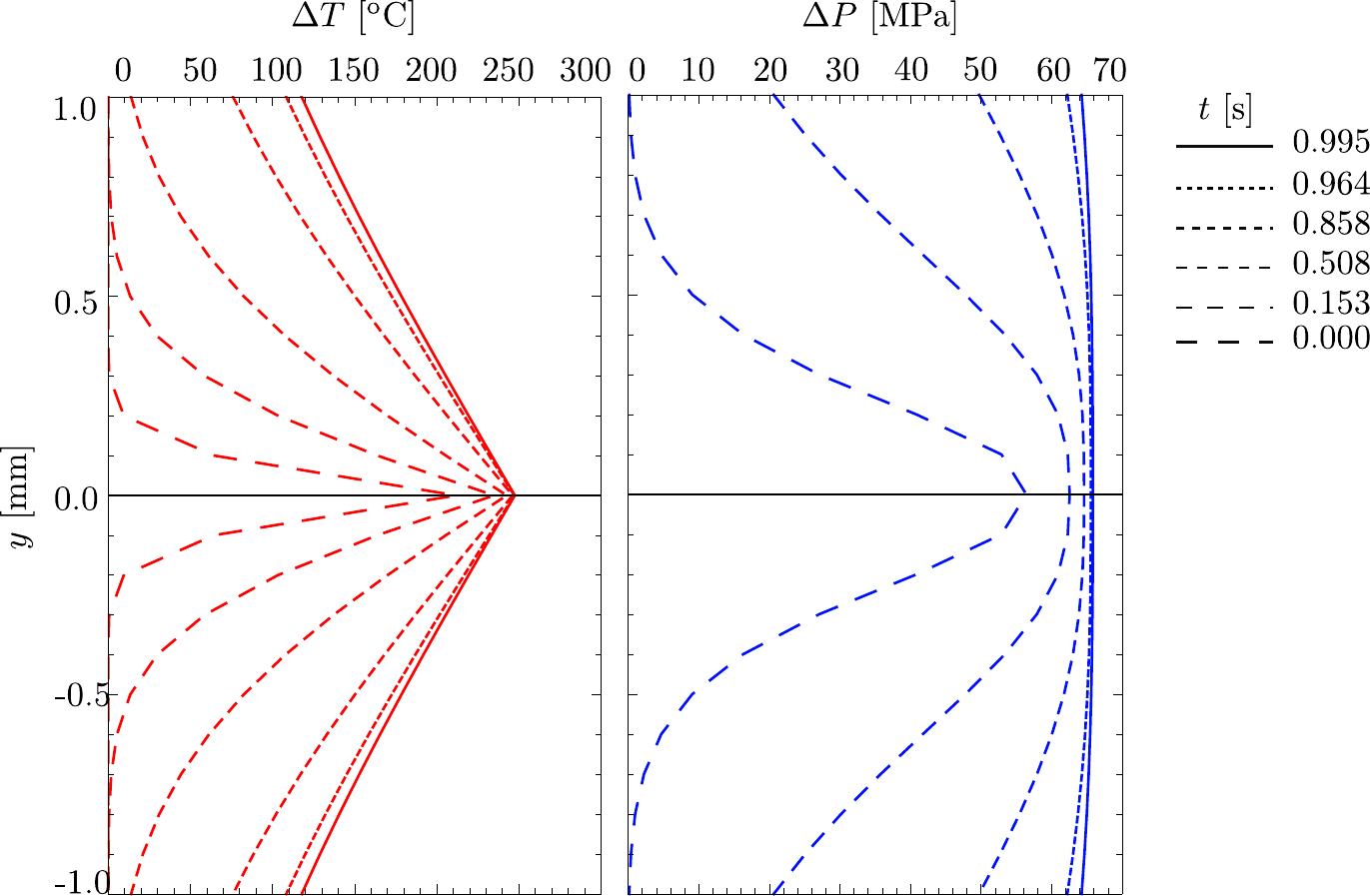}
  \caption{Temperature $\Delta T$ and pore fluid pressure $\Delta P$ fields along the height of the layer for shearing velocity $\dot{\delta}=1$ m/s, at different times during the analysis. The numerical solution is consistent with the analytical observation that the position of $\Delta P_{max}$ coincides with the position of the stationary strain localization.}
  \label{ch: 6 fig: Temp and press fields}  
\end{figure}

\noindent The solutions for the temperature field on an infinite layer under a stationary point source thermal load were first derived in \cite{carslaw1959conduction}. \cite{Mase1987} and \cite{Andrews2005}, present temperature field solutions for stationary distributed thermal loads. Later in \cite{Lee1987} the authors used the above temperature solutions to derive the pressure solution fields $\Delta P(y,t)$ of the coupled pore fluid pressure equation.\\
\newline
\noindent In the work of \cite{Rice2006,Rempel2006} the authors introduce a methodology for the determination of the coupled frictional response of a fault gouge under constant shear rate. The results for the stationary instability on an infinite domain have already been derived in \cite{rice2006heating} for yielding on a mathematical plane, and further expanded in the case of distributed yield in \cite{Rempel2006}. In this case, a closed form analytical solution is possible: $\tau(\delta)=f(\sigma_n-p_0)\exp(\frac{\delta}{L^\star})\erfc(\sqrt{\frac{\delta}{L^\star}})$, where $L^\star=\frac{4}{f^2}\left(\frac{\rho C}{\Lambda}\right)^2\frac{\left(\sqrt{c_{hy}}+\sqrt{c_{th}}\right)^2}{\dot{\delta}}$. The derived solution is recognized as the Hermite polynomial of degree -1.\\
\newline 
\noindent We note that this solution is dependent on the seismic slip rate $\dot{\delta}$ (see dimensions of $L^\star$). The dependence of the fault friction on the seismic slip rate $\dot{\delta}$ (velocity weakening) has been shown in experiments \cite[see][among many others]{Badt2020,Harbord2021,Rempe2020}. In order to demostrate the efficiency of the SCLM method, we use the above analytical solution as a benchmark for comparison. In Figure \ref{ch: 6 fig: frictional_behavior_stationary_unbounded} we present the numerical results of slip on a stationary mathematical plane.
\\
\newline
\noindent To showcase further the accuracy of our results, we present the calculated temperature $\Delta T(y,t)$ and pressure $\Delta P(y,t)$ fields, computed with the method of Gaussian quadrature at the already computed Chebyshev nodes for the time domain, in a uniform spatial grid around the position of strain localization. The results of Figure \ref{ch: 6 fig: Temp and press fields} indicate that at all times the pressure maximum coincides with the position of the strain localization as expected from the analytical solution. This corroborates the accuracy and precision of our results. 
\\
\subsubsection{Stationary strain localization on a bounded domain\label{ch: 6 sec: Stationary_bounded}} 
\begin{figure}[h!]
  \centering
  \includegraphics[width=0.9\linewidth]{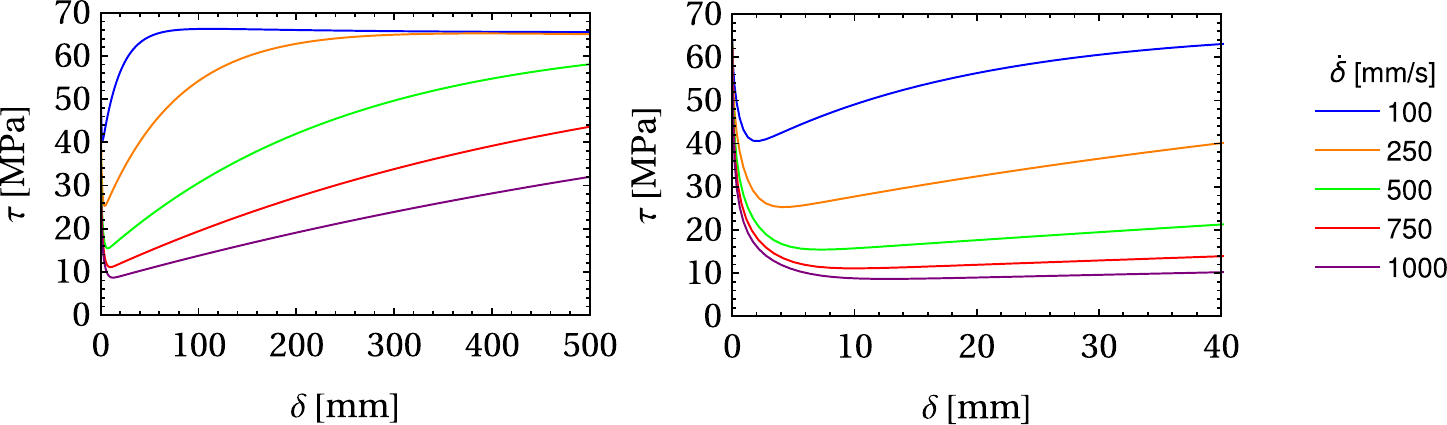}
  \caption{$\tau-\delta$ response of the layer for different slip velocities $\dot{\delta}$ applied. We observe that as the shearing rate increases, the softening behavior becomes more pronounced. For typical values of the seismic slip displacement we note that the effect of the boundaries becomes important. Due to the existence of a steady state the fault recovers all of its strength lost due to thermal pressurization at the beginning of the coseismic slip.}
  \label{ch: 6 fig: frictional_behavior_stationary_bounded}
\end{figure}

\begin{figure}[h!]
  \centering
  \includegraphics[width=0.75\linewidth]{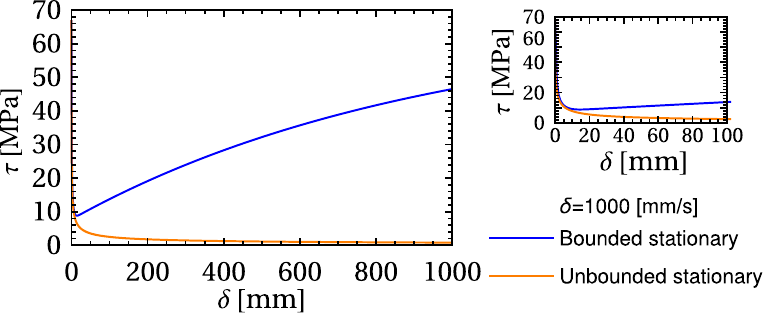}
  \caption{ Comparison of the $\tau-\delta$ response of the layer for an applied slip velocity $\dot{\delta}=1000$ mm/s. We observe that the influence of the boundaries becomes important from the early stages of coseismic slip ($\delta\approx10$ mm). In the bounded case, due to the existence of a steady state the fault tends to recover all of its strength lost to thermal pressurization at the beginning of the phenomenon. namely for a typical value of coseismic slip $\delta=1000$ mm, the fault has recovered more than half of its initial frictional strength.}
  \label{ch: 6 fig: comparison_stationary_bounded_unbounded}
\end{figure}

\begin{figure}[h!]
  \centering
  \includegraphics[width=0.75\linewidth]{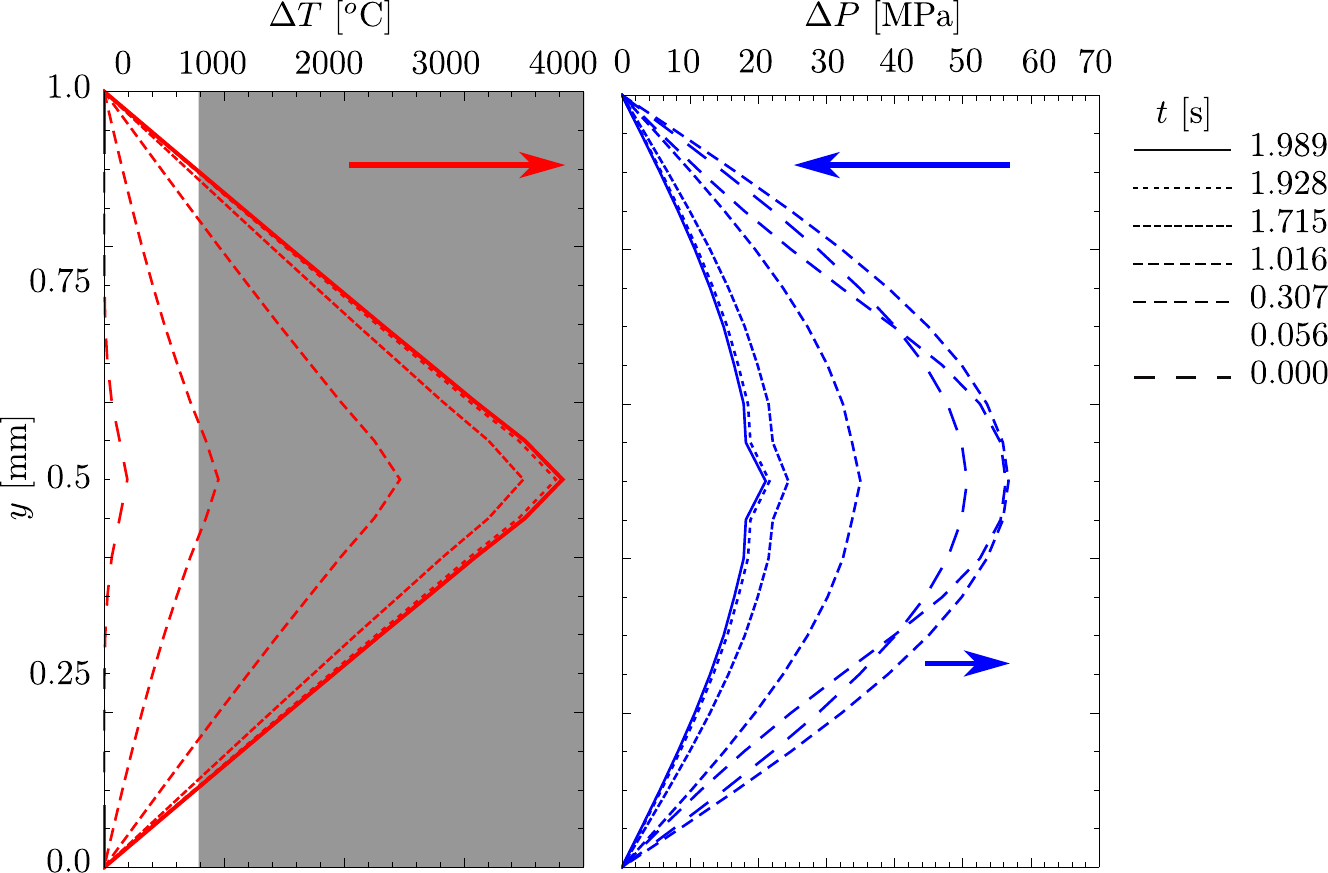}
  \caption{Temperature $\Delta T$ and pore fluid pressure $\Delta P$ fields along the height of the layer for shearing velocity $\dot{\delta}=1$ m/s, at different times during the analysis. The numerical solution is consistent with the analytical observation that the position of $\Delta P_{max}$ coincides with the position of the stationary strain localization. The arrows indicate the evolution course of the maxima of each field. The pressure field initially increases before subsiding when the temperature field progressively reaches steady state and thermal pressurization ceases. The shaded area, indicates a range of temperatures ($\Delta T\geq 800^o$C), that is prohibitively large inside the fault gouge since it corresponds to melting of the gouge material. Moreover, at $\Delta T\geq 600^o$C chemical decomposition of minerals will start to take place inside the gouge, antagonizing the weakening mechanism of thermal pressurization.}
  \label{ch: 6 fig: Temp and press fields bs}
\end{figure}

\noindent When the yielding region (PSZ) is wholly contained on a mathematical plane one might assume that the true boundaries of the fault gouge play little role in the evolution of the phenomenon, simulating the fault gouge region as an infinite layer. However, the validity of this model depends heavily on the pressure and temperature diffusion characteristic times in comparison to the total evolution time of the seismic slip. In essence, the question is: Does the phenomenon evolve so fast that the boundaries do not play a role in the overall frictional response?\\
\newline
\noindent This is a valid question, considering that in experiments and in the majority of the numerical simulations, we need to assign some kind of boundary conditions to the problem in question. We address this question by investigating the case of a stationary strain localization (point thermal source) in the middle of a bounded domain representing the fault gouge, with the linear Volterra integral equation of the second kind \eqref{ch 6: eq: normalized_integral_equation}. We do so by applying the new form of the kernel $G^\star_{X11}(x,x^\prime,t-t^\prime,c_{hy},c_{th})$, which takes into account the boundary conditions of coseismic slip, pressure and temperature discussed in Part I, \cite{Alexstathas2022a}. Namely, the domain of the fault gouge was assumed to have a width of $\text{H}=1$ mm. We remind also that the boundary conditions correspond to an isothermal ($\Delta T(0,t)= \Delta T(\text{H},t)=0$) drained ($\Delta P(0,t)=\Delta P(\text{H},t)=0$) case. \\
\newline
\noindent In order to solve equation \eqref{ch 6: eq: normalized_integral_equation} for the new kind of boundary conditions, we need to derive the new expressions for the Green's function kernel for the thermal diffusion and coupled pore fluid pressure diffusion equations on the bounded domain. The expression for the bounded Green's function kernel under Dirichlet boundary of the heat diffusion equation \eqref{ch 6: eq: normalized_bounded_kernel}, can be found by applying the method of separation of variables according to \cite{cole2010heat}.\\
\newline    
\noindent Equation \eqref{ch 6: eq: normalized_bounded_kernel} is termed the long co-time Green's function kernel. A mathematically equivalent short co-time solution can be constructed making use of the Green's kernel defined for the infinite domain case via the method of images, however, its form is significantly more complicated than equation \eqref{ch 6: eq: normalized_bounded_kernel} and is not convenient for the numerical procedures used in this paper. Namely, the short co-time solution is best suited when studying transient diffusion at the very start of the phenomenon. For fast timescales we don't need a lot of terms for the short co-time series to converge to the expected degree of accuracy. However, for large timescales after the initiation of the phenomenon the large co-time solution converges faster, i.e. using fewer terms in the sum. Furthermore, the form of the large co-time solution has a  simpler form and can be integrated numerically faster, i.e. with less machine operations, than that of the short co-time.\\
\newline
\noindent Next, we need to obtain the Green's function for the coupled pore fluid pressure diffusion equation. This is done by solving the coupled pressure differential equation on the bounded domain, using the method of separation of variables. We note that the two diffusion problems (thermal and coupled pore fluid pressure) are bounded by Dirichlet boundary conditions on the same domain and therefore, their Fourier expansions belong to the same Sturm-Liouville problem. This allows us to express, for the first time in the literature, the Green's function kernel of the coupled temperature diffusion system on a bounded domain due to an impulsive thermal load. Full derivation details are shown in Appendix \ref{Appendix E}, where we prove that the kernel in question can be given in a manner similar to the original expression for the infinite domain case found in \cite{Lee1987}.
\\
\newline
\noindent Next, we apply the kernel of equation \eqref{ch 6: eq: normalized_bounded_kernel} in the equation \eqref{ch 6: eq: normalized_integral_equation}. Using the SCLM method, the values of friction at specific values of time ($t$) and seismic slip displacement (${\delta}$) can be derived for different seismic slip velocities ($\dot{\delta}$). The results of such an analysis are presented in Figure \ref{ch: 6 fig: frictional_behavior_stationary_bounded}.
\\
\newline
\noindent We note here that contrary to the results obtained in the case of the infinite layer in \cite{Rice2006,Rempel2006}, where the frictional response is decreasing monotonously (see also Figures \ref{ch: 6 fig: frictional_behavior_stationary_unbounded},\ref{ch: 6 fig: Temp and press fields}), in the case of the stationary thermal load on a bounded layer the frictional response is eventually influenced by the boundaries of the domain (see Figures 
\ref{ch: 6 fig: comparison_stationary_bounded_unbounded},\ref{ch: 6 fig: Temp and press fields bs}). Since the conditions on the boundaries are constant in time and the frictional source provides heat to the layer at a rate that is bounded by a constant $(\frac{1}{\rho C}\tau(t)\dot{\delta}\leq\frac{1}{\rho C}\tau_0\dot{\delta}=M)$, the temperature field will eventually reach a steady state. This in turn means that at the later stages of the phenomenon the temperature profile will remain constant in time, therefore its rate of change $\frac{\partial \Delta T}{\partial t}$ will become zero. Consequently, the phenomenon of thermal pressurization will cease, leading to rapid pore fluid pressure decrease due to the diffusion at the boundaries. As a result pore fluid pressure will return to its ambient value, and therefore, friction will regain its initial value too. 
\\
\newline
\noindent It is important to note here that as we show in Figure \ref{ch: 6 fig: comparison_stationary_bounded_unbounded}, frictional regain happens well inside the time and coseismic slip margins observed in nature during evolution of the earthquake phenomenon. Of course frictional regain depends on the height of the layer. Namely as the height of the layer increases, the stress drop due to thermal pressurization at the initial stages becomes larger and the fault gouge recovers its frictional strength slower and in later stages of slip. However, the height of the fault gouge H=1 mm corresponds to typical values from fault observations around the globe \cite[see][among others]{myers2004evolution,Rice2006,Sibson2003,sulem2004experimental}. Furthermore, based on the significantly higher hydraulic, and to a lesser extent thermal, diffussivities of the surrounding damaged zone 
\cite[see Part I][]{aydin2000fractures,tanaka2007thermal}, we conclude that the assumption of isothermal drained conditions at the boundaries of the fault gouge as a first approximation, is also justified.We note in particular that for a mature fault gouge, the ratio of the hydraulic  permeability and thermal conductivity of the fault gouge $(^{f})$ to the surrounding damaged zone $(^d)$ lies between $r_{hy}=\frac{k^{f}_{hy}}{k^d_{hy}}=10^2\sim 10^6,\;r_{th}=\frac{c^f_{th}}{c^d_{th}}=1\sim 10$. Therefore, the a priori assumption that an infinite layer describes adequately well the fault gouge during seismic slip should, in our opinion, be revised.
\\
\newline
\noindent Next, we provide in Figure \ref{ch: 6 fig: Temp and press fields bs} the field numerical solutions for the change in temperature and pressure in a bounded domain of height $\text{H}=1$ mm, under constant seismic slip rate $\dot{\delta}=1$ m/s. In the bounded domain, the fields of temperature and pressure will reach the steady state, while the maximum pore fluid pressure coincides with the position of the stationary strain localization. However the steady state reached now is one where full frictional regain takes place. Therefore, the predicted temperature field at the steady state is not applicable, since other weakening mechanisms will take place (e.g. thermal decomposition of minerals will start at 900 $^o$C, see \cite{Sulem2009,Sulem2016}). The role of the boundary conditions at the fault gouge level becomes very important.
\subsection{Traveling mode of strain localization}
\noindent In the available literature \cite{Rice2006,rice2006heating} and the subsequent works \cite{Rempel2006,platt2014stability,rice2014stability} one of the main assumptions is that the principal slip zone (PSZ), which is described by the profile of the plastic strain rate (localized on a mathematical plane or distributed over a wider zone) remains stationed in the same place during shearing of the infinite layer. In this work we depart from this assumption, by assuming that the principal slip zone is traveling inside the fault gouge.\\
\newline
\noindent Two cases will be discussed, the first one discusses the implications of a traveling shear band inside the infinite layer, while the other case focuses on a moving shear band inside the bounded layer. The difference between a stationary and a moving shear band is that in the second case a steady state for the temperature $\Delta T(y,t)$ and pressure $\Delta P(y,t)$ fields is not possible (i.e. their rates of change cannot become zero, $\frac{\partial \Delta T}{\partial t}\neq0,\;\frac{\partial \Delta P}{\partial t}\neq0$ , since the profile of temperature constantly changes due to the thermal load constantly moving around the domain. This ensures that thermal pressurization never ceases. Thus, the value of the residual friction $\tau_{res}$ depends on the fault gouge's thermal and hydraulic properties $(c_{th}, c_{hy})$, the coseismic slip velocity $\dot{\delta}$, and the traveling velocity of the strain localization mode $(v)$. This has serious implications for the frictional response of the layer during shearing. More specifically, as the load does not stay stationary, thermal pressurization does not have enough time to act by increasing the pore fluid pressure. Therefore, according to the Mohr-Coulomb yield criterion, friction does not vanish as in the case of \cite{Rice2006}. Instead friction reaches a residual value $\tau_{res}$ different than zero. This is central for the dissipated energy \cite[see][among others]{Andrews2005,Kanamori2004,Kanamori2006} and the control of the fault transition from steady to unsteady seismic slip. 
\subsubsection{Traveling mode of strain localization in the unbounded domain.\label{ch: 6 sec: Traveling_unbounded}}
\begin{figure}[h]
  \centering
  \includegraphics[width=0.5\linewidth]{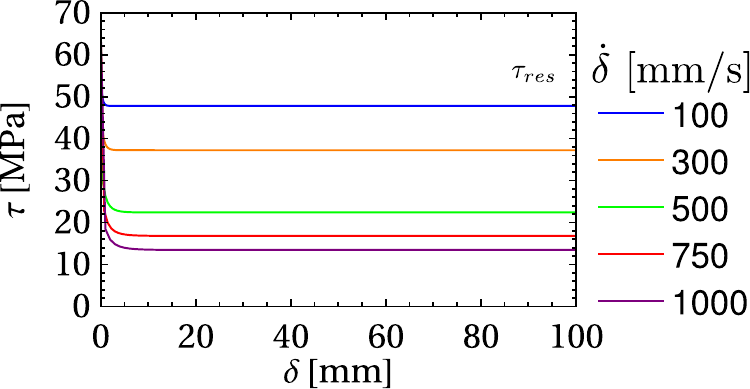}
  \caption{$\tau-\delta$ response of the layer for different slip velocities $\dot{\delta}$ applied. We observe that as the shearing rate increases, the softening behavior becomes more pronounced. Higher seismic slip rates correspond to lower residual values for friction.}
  \label{ch: 6 fig: frictional_behavior_moving_unbounded}  
\end{figure}

\begin{figure}[h]
  \centering
  \includegraphics[width=0.75\linewidth]{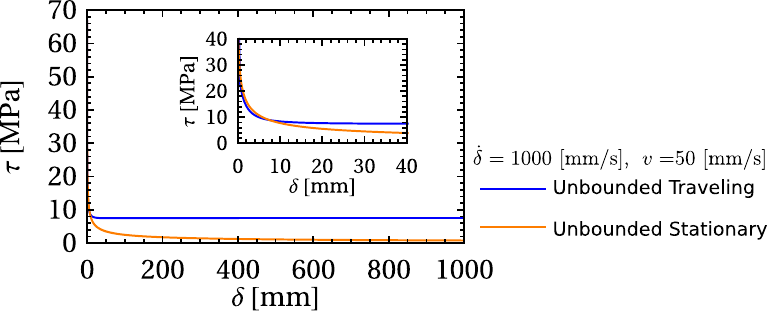}
  \caption{Comparison of the $\tau-\delta$ frictional response between a moving and a stationary strain localization (PSZ) in an unbounded domain. The assumption of a traveling strain localization leads to a plateau of non zero residual friction $\tau_{res}$, contrary to the solution of \protect{\cite{Rice2006}}, which is based on a stationary PSZ.}
  \label{ch: 6 fig: friction_compare_mov_sta_unbounded}  
\end{figure}

\begin{figure}[h!]
  \centering
  \includegraphics[width=0.5\linewidth]{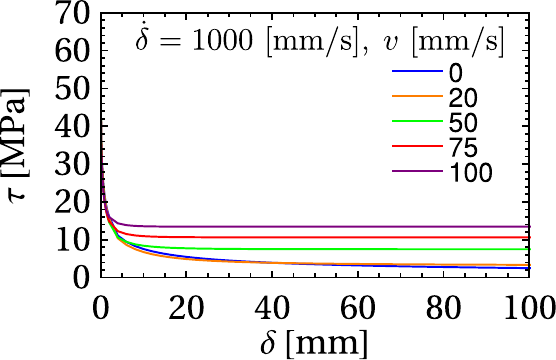}
  \caption{ Frictional response $\tau-\delta$ of the layer for different velocities $v$ of the traveling PSZ. For low traveling velocities the response tends to the behavior of stationary slip on a mathematical plane. As the traveling velocity increases the drop in friction becomes less.}
  \label{ch: 6 fig: frictional_behavior_moving_unbounded_vel}
\end{figure}

\begin{figure}[h!]
  \centering
  \includegraphics[width=0.75\linewidth]{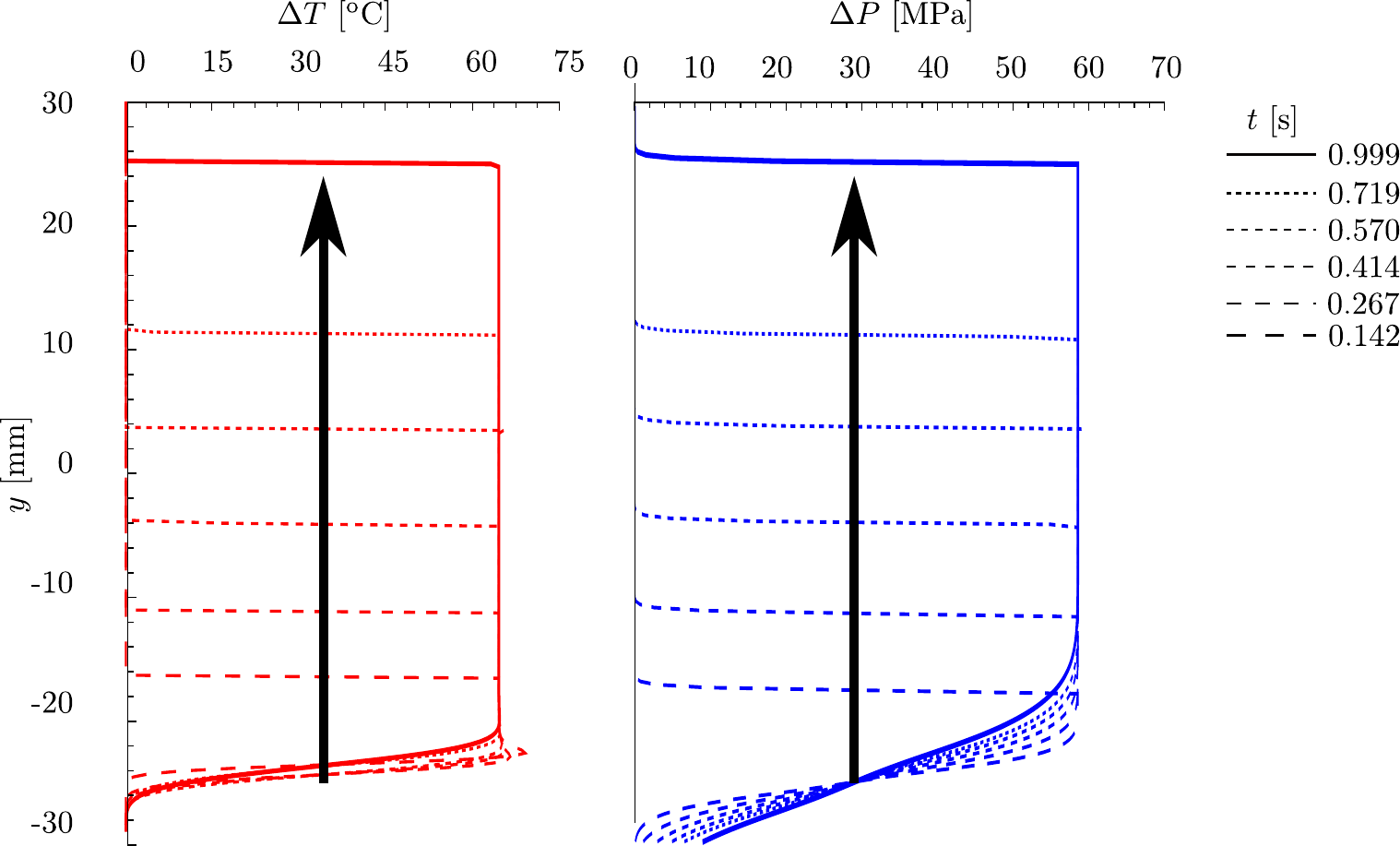}
  \caption{Temperature $\Delta T$ and pore fluid pressure $\Delta P$ fields along the height of the layer for shearing velocity $\dot{\delta}=1$ m/s, at different times during the analysis. The numerical solution is consistent with the analytical observation that the position of $\Delta P_{max}$ coincides with the position of the stationary strain localization.}
  \label{ch: 6 fig: Temp and press fields unbounded moving}  
\end{figure}

\noindent Here we consider the shearing of a fault gouge, whose boundaries are taken at infinity. In what follows, we distinguish between the seismic slip velocity $\dot{\delta}$ and the velocity of the traveling shear band $v(t)$. In Figure \ref{ch: 6 fig: frictional_behavior_moving_unbounded}, we consider the PSZ (moving point heat source) to travel inside the fault gouge with a velocity $v$=50 mm/s. Different values for the rate of coseismic slip parameter $\dot{\delta}$ are taken into account. The shear band velocity $v$ is in agreement with observations from the numerical results of Part I \cite{Alexstathas2022a}. Contrary to the results obtained in the case of a stationary strain localization studied in \cite{Rice2006}, our results indicate the existence of a lower bound in the frictional strength $\tau_{res}$, dependent on the rate of seismic slip $\dot{\delta}$ (see Figure \ref{ch: 6 fig: friction_compare_mov_sta_unbounded}).
\\
\newline
\noindent In Figure \ref{ch: 6 fig: frictional_behavior_moving_unbounded}, we observe that an increase in seismic slip velocity $\dot{\delta}$ leads to a decrease of the frictional plateau. Since the plateau reached in these cases is other than the initial friction value corresponding to the ambient pore fluid pressure, we conclude that thermal pressurization is still present in the model's response. This is true, since the profile of temperature changes continuously, due to the yielding plane moving at a constant velocity $v$. This forces the maximum temperature, $T_{max}$, to move in the same way. Thus, the rate of change of the temperature field $\frac{\partial \Delta T}{\partial t}$, which is the cause of thermal pressurization, does not vanish.\\
\\
\newline
\noindent In Figure \ref{ch: 6 fig: frictional_behavior_moving_unbounded_vel}, we plot the frictional response of the fault for a given seismic slip velocity $\dot{\delta}=1$ m/s treating the shear band velocity $v$ as a parameter. We notice that the slower moving shear bands force the fault to faster and larger frictional strength drops, before they eventually reach a plateau. This is consistent with the observations made in \cite{Rice2006}, where the stationary shear band that presents an infinite negative slope at the start of the slip $\delta$ and tends asymptotically to zero as $\delta$ increases, can be treated as a special case of the model of traveling localization mode as the shear band velocity tends to zero ($v=0$). 
\\
\newline
\noindent In Figure \ref{ch: 6 fig: Temp and press fields unbounded moving}, we present the evolution with time of the temperature $\Delta T(y,t)$ and pressure increase $\Delta P(y,t)$ fields, in the region of the unbounded domain covered by traveling strain localization mode. We note that in this case the traveling localization mode leads to a distribution of the thermal load inside the domain, which -since thermal pressurization remains constant- leads to significantly lower values of temperature inside the domain. We note that the frictional response shown in Figure \ref{ch: 6 fig: frictional_behavior_moving_unbounded} is consistent with the pressure increase $\Delta P(y,t)$ inside the domain, while the temperature and pressure fronts coincide with the prescribed position of the traveling strain localization (thermal load).
\\
\subsubsection{Traveling mode of strain localization in the bounded domain.\label{ch: 6 sec: Traveling_bounded}}
\begin{figure}[h!]
  \centering
  \includegraphics[width=0.75\linewidth]{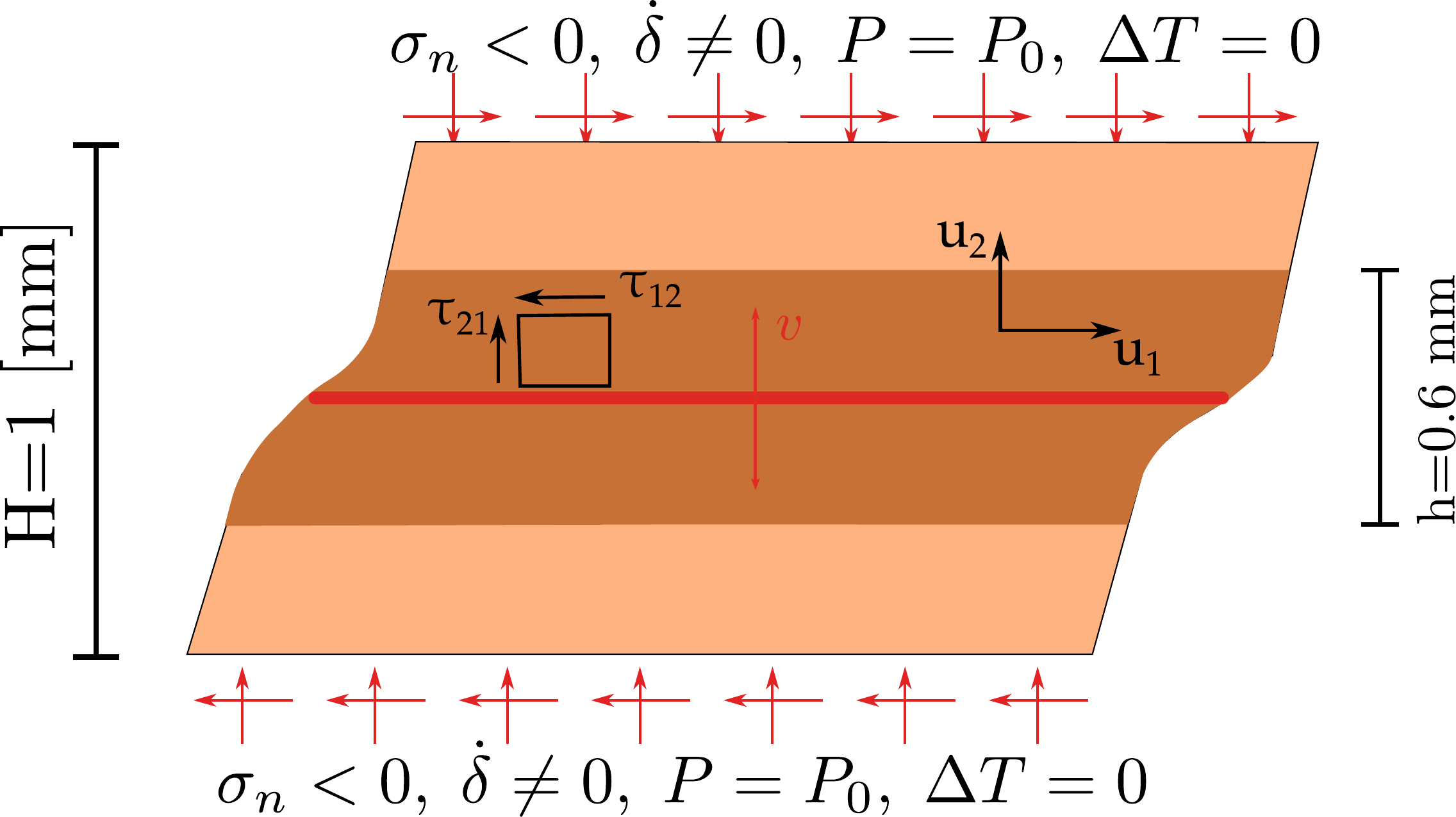}
  \caption{Schematic representation of a fault gouge of height $H=1$ mm, under seismic slip $\delta$. The PSZ - red line- is allowed to travel in a region of thickness h=0.6mm according to the numerical results of Part I, \protect{\cite{Alexstathas2022a}}. The PSZ is moving periodically inside the region $h$ with velocity $v$.}
  \label{ch: 6 fig: explanatory_figure_moving_unbounded}  
\end{figure}
\begin{figure}[h!]
  \centering
  \includegraphics[width=0.75\linewidth]{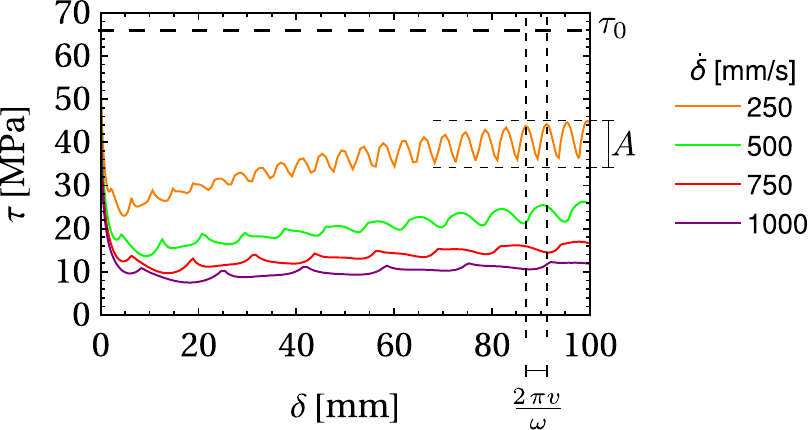}
  \caption{$\tau-\delta$ response of the bounded layer for different slip velocities $\dot{\delta}$ applied. A periodic traveling localization mode is applied. We observe that as the shearing rate increases, the softening behavior becomes more pronounced. For typical values of the seismic slip displacement we note that the effect of the boundaries becomes important. As the periodic traveling localization mode is constantly moving, a steady state is not possible. This means that the friction presents an oscillating residual value lower than the fully recovered value of the stationary bounded case. }
  \label{ch: 6 fig: frictional_behavior_moving_bounded}  
\end{figure}
\begin{figure}[h!]
  \centering
  \includegraphics[width=0.9\linewidth]{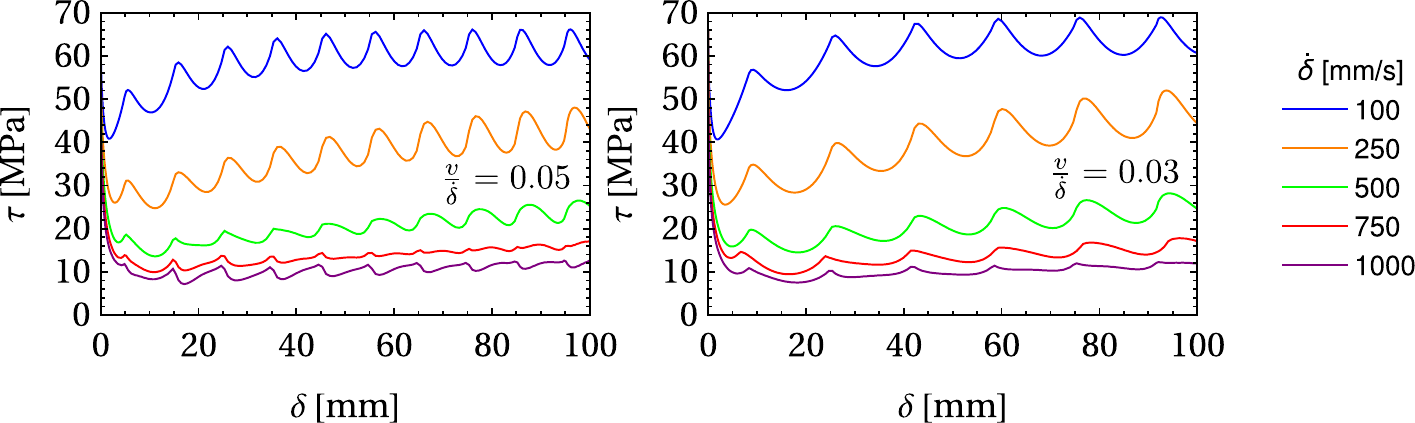}
  \caption{$\tau-\delta$ response of the bounded layer for different ratios of strain localization velocities $v$ to coseismic slip rates $\dot{\delta}$ applied $(\frac{c}{\dot{d}})$. We note that for the same rate the period of oscillation remains the same. The period of oscillations is depends on the height of the layer $H$ and the velocty of the strain localization.}
  \label{ch: 6 fig: frictional_behavior_moving_bounded_vd_compare}  
\end{figure}
\begin{figure}[h!]
  \centering
  \includegraphics[width=0.75\linewidth]{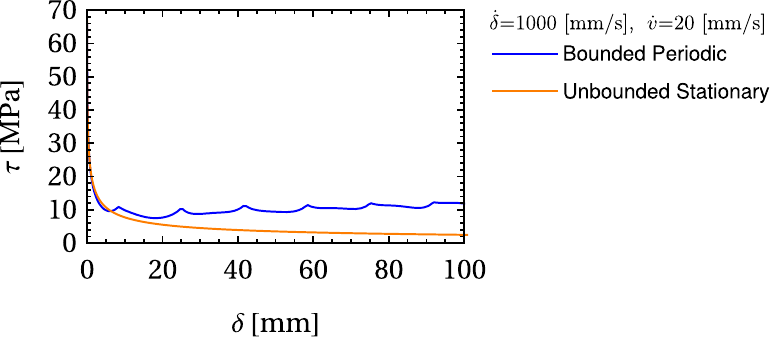}
  \caption{Comparison of the $\tau-\delta$ frictional response between a moving periodic strain localization on a bounded domain and a stationary strain localization (PSZ) on an unbounded domain. The influence of the boundary conditions is noticeable from the initial stages of the coseismic slip ($\delta\approx 10$ mm).}
  \label{ch: 6 fig: frictional_behavior_moving_per_bounded_sta_unbounded}  
\end{figure}
\begin{figure}[h!]
  \centering
  \includegraphics[width=0.75\linewidth]{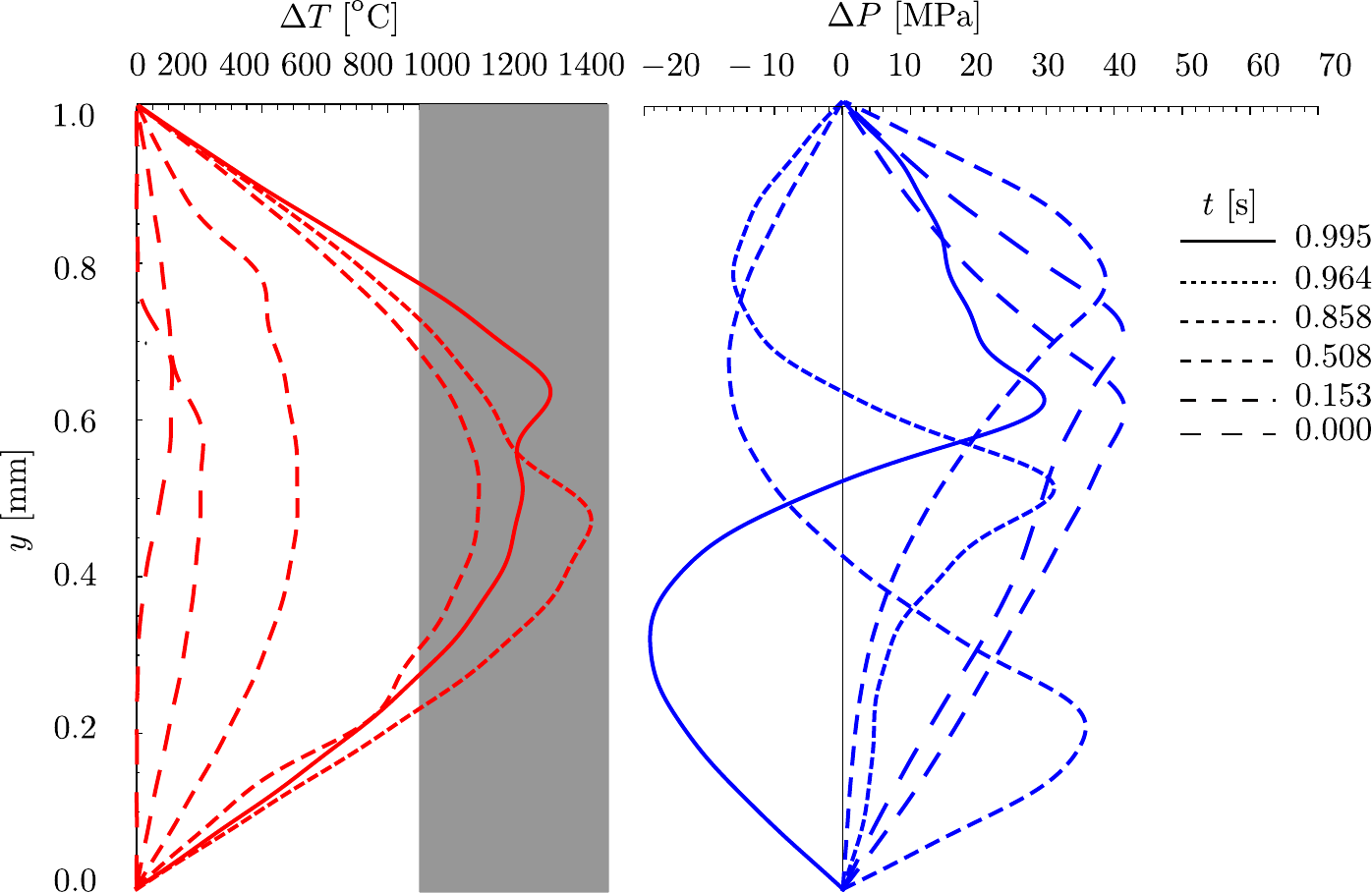}
  \caption{Temperature $\Delta T$ and pore fluid pressure $\Delta P$ fields along the height of the layer for shearing velocity $\dot{\delta}=1$ m/s, at different times during the analysis. 
  Because of the thermal load moving inside the domain closer to the sinks at the boundaries, temperature reaches markedly smaller values than in the stationary case. We note that the change in the pressure field presents negative values leading to regions of smaller pressure than the initial $P_0$ ($P(y,t)=P_0+\Delta P(y,t)$). Thiscoincides with the numerical analyses presented in Part I, \cite{Alexstathas2022a}.}
  \label{ch: 6 fig: fields_moving_bounded}  
\end{figure}

\noindent In this section we investigate the frictional response of the layer of height $\text{H}=1$ mm, when the plastic strain localization (PSZ) travels inside a predefined region with a width $\text{h}=0.6$ mm as shown in Figure \ref{ch: 6 fig: explanatory_figure_moving_unbounded}. This region has the same width as the width of the plastified region predicted by our numerical model in Part I, \cite{Alexstathas2022a} see Figure \ref{Part I-ch: 5 fig: l_dot_gamma_final_1000-T_p_final_1000}). Based on the numerical results of Part I, we apply a periodic mode of traveling strain localization, with a constant velocity $v=30$ mm/s. We prescribe the trajectory of the yielding plane, whose position $u(t)$ is given by a triangle pulse train:
\begin{align}
u(t)=\frac{\text{H}}{2}+\frac{\text{h}}{2\text{H}}Tr(v t),
\end{align}
\noindent where $\text{H}$ is the height of the layer, $\text{h}$ is the width of the plastified region, $v$ is the velocity of the strain localization and $Tr(\cdot)$ is the triangle wave periodic function. The period is given by $T=\frac{2\text{h}}{v}$. 
The resulting linear Volterra integral equations of the second kind is solved numerically by making use of the spectral collocation method in section \ref{ch: 6 sec: section 3}. \\
\newline
\noindent We observe in Figure \ref{ch: 6 fig: frictional_behavior_moving_bounded} that as the shearing rate increases, the softening behavior becomes more pronounced. For typical values of the seismic slip displacement we note that the effect of the boundaries becomes important. The frictional response presents oscillations due to the periodic movement of the strain localization. Since the strain localization is constantly moving, a steady state is not possible for the fields of temperature and pressure ($\frac{\partial \Delta T}{\partial t}\neq 0\to\frac{\partial \Delta P}{\partial t}\neq 0$). This means that the friction presents a residual value, $\tau_{res}$, which is lower than the fully recovered value of the stationary bounded case.
\\
\noindent Assuming the material parameters $c_{th},c_{hy}$ and the height of the layer $\text{H}$ constant, characteristics such us the oscillations amplitude $A$, circular frequency $\omega$ and the residual value of friction $\tau_{res}$ are controlled by three parameters, the thickness of the prescribed region the PSZ is allowed to travel inside the layer, $h$, the velocity of the traveling PSZ, $v$, and the seismic slip rate applied at the fault gouge, $\dot{\delta}$. \\
\newline
\noindent In Figure \ref{ch: 6 fig: frictional_behavior_moving_bounded_vd_compare}, we investigate the influence of the shearing velocity $\dot{\delta}$, the velocity of the traveling shear band $v$ on the frictional response of a fault gouge with height $\text{H}=1$ mm. We note that the period of oscillations in the frictional response depends on the velocity with which the shear band travels inside the fault gouge. For the range of applied traveling shear band velocities $30-50$ mm/s the minima and maxima of the frictional response $\tau-\delta$ are not affected. 
\newline
\noindent In Figure \ref{ch: 6 fig: frictional_behavior_moving_per_bounded_sta_unbounded}, we present a comparison between the friction developed during shearing of a bounded fault gouge and the model of slip on a stationary mathematical plane presented in section \ref{ch: 6 sec: Rice_proc_inf_layer} and in \cite{Rice2006}. In the bounded fault gouge, the seismic slip velocity is given by $\dot{\delta}=1000$ mm/s. We further consider the shear band to travel with a velocity $v=30$ mm/s inside a predefined region of height $\text{h}=0.6$ mm. We note that the two responses differ. 
 The periodic movement of the yielding plane (thermal load) inside the layer leads to frictional oscillations. This happens because the yielding plane moves towards the isothermal drained boundaries that function as heat and pressure sinks. Namely, the crests of the oscillations correspond to the time instance the load approaches the fault gouge boundaries, while troughs correspond to the time the PSZ is closer to the middle of the layer. We note here that the average friction inside the layer, $\tau_{ave}$, is increasing due to the diffusion of pressure and temperature at the boundaries of the fault gouge. We note also that the oscillatory movement of the fault gouge moves excess heat and pressure towards the boundaries of the fault gouge leading to a ventilation phenomenon that further enhances the recovery of frictional strength. It is likely that removing the invariance along the slip direction would lead to vortices and other convective phenomena inside the layer \cite[see][]{griffani2013rotational,miller2013eddy,rognon2015circulation}. However, 2D and 3D phenomena inside the fault gouge are not explored here.
\\
\newline
\noindent The results obtained in Figures \ref{ch: 6 fig: frictional_behavior_moving_bounded},\ref{ch: 6 fig: frictional_behavior_moving_bounded_vd_compare}, \ref{ch: 6 fig: frictional_behavior_moving_per_bounded_sta_unbounded}, present a qualitative agreement with those of Part I \cite{Alexstathas2022a}. The difference in the values is due to the assumption of a Dirac load in this paper, in order to preserve the equilibrium inside the band. Assuming a distribution of the yielding rate $\dot{\gamma}^p$ that is not singular while respecting the equilibrium conditions along the layer -as it is the case for the Cosserat continuum- would allow for higher minima in the frictional response, because of the distributed thermal load over the finite thickness of the yielding region. This leads to more efficient diffusion at the initial stages of thermal pressurization. \\
\newline
\noindent In Figure \ref{ch: 6 fig: fields_moving_bounded} we present the fields of temperature $\Delta T(y,t)$ and pore fluid pressure increase $\Delta P(y,t)$ during shearing of the bounded fault gouge, with coseismic slip rate $\dot{\delta}=1$ m/s, assuming a traveling mode of strain localization traveling with a velocity of $v=30$ mm/s. We note that along the bounded fault gouge, the pore fluid pressure increase might become negative. This is acceptable as long as the total pore fluid pressure doesn't become negative ($\Delta P(y,t)>-P_0$). This is a characteristic that also exists in our fully nonlinear numerical analyses on the bounded domain (see \cite{Alexstathas2022a}, Figure \ref{Part I-ch: 5 fig: l_dot_gamma_final_1000-T_p_final_1000}). 
\\
\section{Conclusions} 
\noindent In this paper a series of numerical results have been obtained for the coupled thermal and pore fluid pressure diffusion equations. We follow the methodology developed in \cite{Rice2006,Rempel2006}, and we expand it to the cases of bounded domains and moving thermal loads resulting from traveling (flutter) instabilities on a Cauchy continuum \cite[see][]{Rice2006,Benallal2003,benallal2005localization,Rice2014,Platt2014}.\\
\newline
\noindent To handle the integral differential equations the SCLM method was applied \cite[see][]{Elnagar1996,tang2008spectral}. The method can handle the weakly singular kernels that appear in the unbounded case and the stationary thermal load on the bounded case. The method can also generalize to the case of a periodic traveling strain localization inside the bounded domain, which is in accordance with the numerical results of Part I, \cite{Alexstathas2022a}.\\
\newline 
\noindent It is found that contrary to the case of a stationary thermal load on an unbounded domain described in \cite{Rice2006}, taking into account the existence of the boundary conditions at the edges of the fault gouge plays an important role at the frictional evolution of the fault for a range of values of the seismic slip velocities commonly observed during earthquake events. Namely, for a seismic slip $\delta$ of 1 m under a seismic slip velocity $\dot{\delta}$=1 m/s, the influence of the boundaries becomes important after the first 0.1 m of slip. It is shown that under the influence of homogeneous Dirichlet conditions on the bounded domain, a steady state is reached for the temperature field, which in turn implies that the effects of thermal pressurization progressively attenuate until it completely ceases. In this case the temperature rise inside the fault gouge is well above the melting point of the fault gouge material. The apparent scarcity of pseudodactilites and absence of widespread melting observations in faults (see \cite{Brantut2008,kanamori2004physics,Rice2006}), however, indicates that other possible frictional weakening mechanisms will become prevalent, such as chemical decomposition of minerals \cite[see][]{Sulem2009}.\\
\newline
\noindent Furthermore, the effects of a moving thermal load corresponding to a traveling strain localization (flutter instability) inside the fault gouge, were examined under both unbounded and bounded boundary conditions. In both cases, traveling strain localization mode showed the existence of a plateau in the frictional strength of the fault, $\tau_{res}$ (see Figures \ref{ch: 6 fig: frictional_behavior_moving_unbounded}, \ref{ch: 6 fig: frictional_behavior_moving_bounded}).\\
\newline
\noindent In the case of the traveling load on the unbounded domain, the fact that the load changes its position constantly leads to a non zero change of the temperature field ($\frac{\partial \Delta T(x,t)}{\partial t}\neq 0$) and constant influence of the pore fluid pressure profile by the thermal pressurization term. Moreover, because the thermal load changes its position, temperature does not have time to accumulate in one point and provoke a pressure increase that eliminates fault friction. Instead fault friction reaches a non-zero plateau (see Figure \ref{ch: 6 fig: frictional_behavior_moving_unbounded}). This is an important result since it directly influences the dissipation energy produced during seismic slip.\\
\newline
\noindent Moreover, we examined the influence of the velocity of the strain localization (moving thermal load) in the frictional evolution. Based on our analyses, we established that the faster traveling shear bands have a smoother stress drop at the first stages of the analysis and they reach a higher plateau of frictional strength, see Figure \ref{ch: 6 fig: frictional_behavior_moving_unbounded_vel}. When the velocity of the shear band tends to zero we retrieve the solution described in \cite{Rice2006}, as expected.\\
\newline
\noindent Next, a traveling instability was applied into a bounded domain with homogeneous Dirichlet boundary conditions. Again the results show that the frictional strength of the fault reaches a plateau and is not fully recovered as in the case of a stationary instability (see Figure \ref{ch: 6 fig: frictional_behavior_moving_bounded}). The reason is the change of the position of the thermal load during the analysis and the subsequent change of the temperature profile, leading to a non attenuating thermal pressurization phenomenon. Again the plateau reached, differs based on the traveling velocity of the shear band $v$, which ranges in the order of $20\sim 50$ mm/s according to the numerical analyses of Part I, \cite{Alexstathas2022a}. In this case, it is shown that in contrast to the case of a stationary thermal load on the bounded domain, the fault never recovers entirely its frictional strength since the effects of thermal pressurization never cease.\\
\newline 
\noindent The results presented above clearly show a strong dependence of the fault's frictional behavior in both the fault gouge boundary conditions and the strain localization mode (traveling or stationary PSZ) introduced into the medium. These results can be used as a preliminary model in order to evaluate qualitatively the results obtained by numerical analyses taking into account the microstructure of the fault gouge material, where discerning between the effects of the different mechanisms affecting the frictional response of a fault undergoing thermal pressurization is more involved. The results of the fully non-linear numerical analyses with the Cosserat micromorphic continuum of Part I agree qualitatively with the results from the linear model of this paper. This indicates that the driving cause behind the obtained results is the diffusion from the thermal and hydraulic couplings. The microstructure follows to a lesser extend. Its use in the solution of the BVP presented in Part I (see \cite{Alexstathas2022a}), is required in order for the dissipation and the meta-stable frictional response of the fault gouge to be calculated correctly excluding mesh dependency from the numerical results. 
\\
\newline
\noindent In conclusion, our results show that for typical values of seismic slip $\delta$ and seismic slip velocity $\dot{\delta}$, the effects of the boundaries of the fault gouge cannot be ignored. This means that those effects need to be accounted in both numerical analyses and laboratory experiments. The influence of different kind of boundary conditions needs to be studied. The introduction of a traveling (flutter-type) strain localization mode is an important aspect of our model. Its presence increases the frequency content of the earthquake and it prevents the bounded fault gouge from fully recovering its frictional shear strength due to the diffusion at the boundaries. Furthermore, it contributes in keeping the temperatures inside the fault gouge smaller than in the stationary cases. The existence of oscillations and the reduction of the peak residual frictional strength are also important in understanding the transition form a stable to unstable seismic slip and subsequent fault nucleation \cite[see][among others]{Rempel2006,Rice1973a,Rice2006,viesca2015ubiquitous}. Furthermore, the existence of non zero upper and lower bounds in the fault's frictional behavior ($\tau_{min},\tau_{res}$), has serious implications for any attempt in controling the transition from stable (aseismic) to unstable (coseismic) slip 
\cite[see][]{Stefanou2019,Stefanou2020,tzortzopoulos2021Thesis}.
\section*{Acknowledgments}
\noindent The authors would like to acknowledge the support of the European Research Council (ERC) under
the European Union’s Horizon 2020 research and innovation program (Grant agreement no. 757848
CoQuake).
\let\clearpage\relax
\appendix
\section{ The coupled Thermo-hydraulic problem and is solution \label{Appendix A}}
\subsection{Problem description}
\noindent We have already discussed that we are interested in the limiting case where the position and profile of the PSZ can be prescribed inside the fault gouge. Knowing the form of the profile of the shear plastic strain-rate in $\dot{\gamma}^p(y,t)$ (equation \eqref{ch: 6 plastic_strain_rate_prf}), the two way coupled problem in the form of temperature and pressure diffusion equations is given by:
\begin{itemize}
\item[•] Heat diffusion BVP:
\begin{align}
&\frac{\partial \Delta T}{\partial t}=c_{th}\frac{\partial \Delta T}{\partial y^2}+\frac{1}{\rho C}\tau(t)V\delta_{\text{Dirac}}(y-u(t)),\nonumber\\
&\Delta T\big{\|}_{y=0}=\Delta T\big{\|}_{y=\text{H}}=0,\nonumber\\
&\Delta T(y,0)=0, \label{ch: 6 Temp_known_gamma}
\end{align}
\noindent where $T(y,t)$ is the unknown change in temperature in the fault gouge layer of height $\text{H}$. The fault gouge is considered to be in isothermal boundary conditions during shearing. The coupled pressure problem is given by:
\item[•] Pressure diffusion BVP, in its Homogeneous form:
\begin{align}
&\frac{\partial \Delta P}{\partial t}=c_{hy}\frac{\partial \Delta P}{\partial y^2}+\Lambda\frac{\partial \Delta T}{\partial t},\nonumber\\
&\Delta P\big{\|}_{y=0}=\Delta P\big{\|}_{y=H}= \Delta P(y,t)-\Delta P(y,0) =0,\nonumber\\
&P(y,0)=P_0, \label{ch: 6 Press_known_gamma}
\end{align}
\noindent where $\Delta P(y,t)$ is the unknown pressure difference between the fault gouge layer and the boundaries, while $P_0=P(y,0)$ is the initial pore fluid pressure, that is kept constant at the boundaries of the fault gouge (drained boundary conditions). 
\end{itemize}
We note that the above formulations are also valid in the case of an unbounded domain considering $\text{H}\to\pm\infty$. The pressure problem affects also the temperature BVP through the value of shear stress (fault friction), $\tau(t)$, in the yielding region. According to the Mohr-Coulomb yield criterion, subtracting the initial ambient pore fluid pressure $P_0$ we get:
\begin{align}
\tau(t)=f(\sigma_n-p_0)-f\Delta P_{max}(t). \label{ch: 6 material_eq_known_gamma}
\end{align}
\noindent We note here that once we know the form of the plastic strain-rate profile $\dot{\gamma}^p(y,t)$ as in equation \eqref{ch: 6 Temp_known_gamma} the only unknown is the fault friction $\tau(t)$. We can find the solution of the temperature equation $T(y,t)$ in terms of the unknown fault friction $\tau(t)$ and replace into the pressure equation, which can then also be described as an unknown function of friction. Finally, we can define the value of fault friction $\tau(t)$ by inserting the pressure increase solution $\Delta P(y,t)$ into the material equation \eqref{ch: 6 material_eq_known_gamma} and solving for $\tau(t)$. The above equations have constant coefficients and since the loading is prescribed (based on the unknown $\tau(t)$), the system has been transformed to a one-way coupled set of linear differential 1D diffusion equations of the form:
\begin{align}
&\frac{\partial q(y,t)}{\partial t} = c \frac{\partial^2 q(y,t)}{\partial x^2}+\frac{1}{k}g(y,t),\nonumber\\
&a_1\frac{\partial u}{\partial n_1}\big{\|}_{y=0}+b_{1}q\big{\|}_{y=0}(t)=f_1(t),\;{t>0},\nonumber\\
&a_2\frac{\partial q}{\partial n_2}\big{\|}_{y=H}+b_{2}u\big{\|}_{y=H}(t)=f_2(t),\;{t>0},\nonumber\\
&q(y,0) = I(y),
\end{align}
\noindent where $q(y,t)$ is the unknown function (e.g. the temperature $T(y,t)$), $f_i,\;i=1,2$ are the values of the general Robin boundary conditions with coefficients ($a_i,\;b_i,\;{i=1,2}$), $I(y)$ is the initial condition and $g(y,t)$ is the loading function (here related to frictional dissipation). We denote by $c$ the diffusivity and by $k$ the conductivity of the material.
\subsection{Fundamental solution \label{ch: 6 subsec_2.2}}
\noindent We can find the solution to the above BVP by application of the Green's theorem, which for the general diffusion case in 1D reads (see \cite{cole2010heat}):
\begin{align}
\begin{aligned}
q(y,t)=&\int_{0}^{\text{H}}G(y,t;y^\prime,0,c)I(y^\prime)dy^\prime+\frac{c}{k}\int_0^t\int_{0}^{\text{H}}g({y^\prime,t^\prime})G(y,t;y^\prime,t^\prime,c)dy^\prime dt^\prime\\
&+c\int_0^t\sum_{i=1}^2\left[\frac{f_i(t^\prime)}{a_i}G(y,t;y^\prime_i,t^\prime,c)\right]dt^\prime-\alpha\int_0^t\sum_{i=1}^2\left[f_i(t^\prime)\frac{G(y_i,t;y^\prime,t^\prime,c)}{n_i^\prime}\bigg{\|}_{y^\prime=y_i}\right]dt^\prime,\label{ch: 6 eq: Green's solution }
\end{aligned}
\end{align}
\noindent where $G(y,t;y^\prime,t^\prime,c)$ is the appropriate Green's function. The first two terms correspond to the initial condition $I(y,0)$ and the loading term $g(y,t)$ respectively. The terms $\alpha ,k$ represent the diffusivity and the conductivity of the unknown quantity $q(y,t)$ respectively. The third term is important for non homogeneous Neumann and Robin boundary conditions while the fourth term refers to non homogeneous Dirichlet boundary conditions. In what follows the last two terms in equation \eqref{ch: 6 eq: Green's solution } are omitted due to the existence of homogeneous Dirichlet boundary conditions in the problems of temperature and pressure difference diffusion at hand.\\
\newline
\noindent Applying the solution in terms of the Green's function \eqref{ch: 6 eq: Green's solution } to problems \eqref{ch: 6 Temp_known_gamma},\eqref{ch: 6 Press_known_gamma} we obtain the solution in terms of the Green's function specific to each diffusion problem.
\begin{align}
&\begin{aligned}
\Delta T(y,t)&=\frac{c_{th}}{k_T}\int_0^t\int_{-\infty}^{\infty}g_T(y^\prime,t^\prime)G(y,t;y^\prime,t^\prime,c_{th})dy^\prime dt^\prime,\\
\end{aligned}\\
&\begin{aligned}
\Delta P(y,t)&= \frac{c_{hy}}{k_H }\int_0^t\int_{-\infty}^{\infty}g_H(y^\prime,t^\prime)G(y,t;y^\prime,t^\prime,c_{hy})dy^\prime dt^\prime,\\
\end{aligned}
\end{align}
\noindent where $c_{th},k_T$ are the thermal diffusivity, conductivity pair and $c_{hy},k_H$
are their hydraulic counterparts. Similarly $(g_{T},g_{H})$ are the loading functions, while $G(y,t;y^\prime,t^\prime,c)$ is the Green's function kernel for the thermal ($c=c_{th}$) and pressure ($c=c_{hy}$) diffusion problems respectively.\\
\newline
\noindent In the case of the coupled pressure problem \eqref{ch: 6 Press_known_gamma} with the temperature as a loading function, we are interested in rewriting the system's response with the help of the dissipative loading $(\frac{1}{\rho C}\tau(t)\dot{\gamma^p})$ of the temperature equation \eqref{ch: 6 Temp_known_gamma}. This way we can connect the pressure response $P(y,t)$ to the fault friction $\tau(t)$ which is the main unknown. We can do this by replacing in the expression of $T(y,t)$ in the pressure diffusion equation \eqref{ch: 6 Press_known_gamma} the temperature impulse response of equation \eqref{ch: 6 Temp_known_gamma} due to a impulsive (Dirac) thermal load. This way the response obtained from the pressure diffusion equation is a Green's function kernel that contains the influence of an impulse thermal load (see Appendix \ref{Appendix E} for detailed derivation in the cases of 1) a bounded domain for a stationary impulsive thermal load and 2) an unbounded domain subjected to a moving impulsive thermal load). The pressure solution can then be written as:
\begin{align}
\Delta P(y,t)=P(y,t)-P_0&= \frac{\Lambda\dot{\delta}}{\rho C(c_{hy}-c_{th})}\int_0^t\int_{-\infty}^{\infty}g_T(y^\prime,t^\prime)G^\star(y,t;y^\prime,t^\prime,c_{hy},c_{th})dy^\prime dt^\prime,\label{ch: 6 eq: pressure_sol_mod}
\end{align}
\noindent where $G^\star(y,t;y^\prime,t^\prime,c_{hy},c_{th})$ is the Green's function kernel of the pressure equation \eqref{ch: 6 Press_known_gamma} containing the influence of an impulse thermal load from the temperature equation \eqref{ch: 6 Temp_known_gamma}.\\
\newline
\noindent Having found the pressure solution $P(y,t)$ as a function of $g_{T}$ we can then replace \eqref{ch: 6 eq: pressure_sol_mod} into the material description equation \eqref{ch: 6 material_eq_known_gamma}. For the case of 1D shear $\tau$ under constant normal load $\sigma_n$ that we will consider throughout this paper, the material law is transformed into the integral equation:
\begin{align}
\tau(t) = f(\sigma_n-P_0)-C\int_0^t\int_{-\infty}^{\infty}g_T(y^\prime,t^\prime)G^\star(y,t;y^\prime,t^\prime,c_{hy},c_{th})dy^\prime dt^\prime, \label{ch: 6 eq: integral equation}
\end{align}
\noindent where $C=\frac{f\Lambda\dot{\delta}}{\rho C(c_{hy}-c_{th})}$. Due to the concentrated nature of the thermal load (Dirac distribution), the integral equation \eqref{ch: 6 eq: integral equation} can be brought to its final form:
\begin{align}
\tau(t) = f(\sigma_n-p_0)-C\int_0^t\tau(t^\prime)G^\star(y,t;y^\prime,t^\prime,c_{hy},c_{th})\big{\|}_{y=y^\prime(t^\prime)} dt^\prime \label{ch: 6 eq: integral equation1}
\end{align}
\noindent The above integral equation is a linear Volterra integral equation of the second kind \cite{wazwaz2011linear}. We note here that this equation is valid only at the position of the yielding plane which has to coincide with the position of the maximum pressure inside the layer ($y=y^\prime(t^\prime)$). This has been proven to hold true for the cases present on the unbounded domain (see Appendix \ref{Appendix I}). In the case of a bounded fault gouge under the influence of a traveling PSZ (thermal load) this is true only in the regions again from the boundary. Nevertheless, the difference between the position of the traveling thermal load and that of the $P_{max}$ is small (see also Figure \ref{ch: 6 fig: fields_moving_bounded}). 
\section{Derivation of the coupled pore fluid pressure diffusion kernel. \label{Appendix E}}
\noindent In this appendix we derive the coupled pore fluid pressure diffusion kernel for the cases of a bounded domain subjected to a stationary Dirac load and an unbounded domain under a moving Dirac load. Our procedure follows the discussion in \cite{Lee1987} where the same problem was solved for a stationary Dirac thermal load on an unbounded domain.
\subsection{Stationary thermal load, coupled pore fluid pressure Green's kernel for a bounded domain.}
\noindent In the case of the bounded domain we proceed by applying the method of separation of variables and then expanding the solution to a Fourier series. We note here that the coupled system of pressure and temperature diffusion equations have the same form of linear partial differential operators and boundary conditions and therefore their solution belongs to the same space of Sturn-Liouville problems. In essence the two solutions have the same eigenfunctions. In the case of the bounded domain the Temperature diffusion equation has the solution given in \cite{cole2010heat}:
\begin{align}
\Delta T(y,t) = \sum_{n=1}^\infty\frac{2}{\text{H} \rho C}\int_0^{t}\int_{-\infty}^\infty g(y^\prime, t^\prime)\exp{\left[-\lambda^2 c_{th}(t-t^\prime)\right]}\sin{\left(\lambda_n x\right)}\sin{\left(\lambda_n y^\prime\right)}dy^\prime dt^\prime,
\end{align} 
\noindent where $\lambda_n$ is the Sturm-Liouvile eigenfunction coefficient, $\lambda_n=\frac{n\pi}{\text{H}}$, $\text{H}$ is the length of the bounded domain. The eigencondition for the homogeneous Dirichlet boundary condtitions are given by:
\begin{align}
\sin\left({\frac{n\pi}{\text{H}}\text{H}}\right)=0,\; \lambda_n=\frac{n\pi}{\text{H}},\; n=1,2,...
\end{align}
\noindent We note here that the homogeneous pressure diffusion partial deifferential equation on the above bounded domain has the same boundary conditions. Therefore, the pore fluid pressure solution can be written with the same eigenfunctions as above. Replacing the pore fluid pressure eigenfunction expansion $\Delta P(x,t)=P(y,t)-P_0=\sum_{i=n}^\infty \tilde{p}_n\sin{\frac{n\pi x}{\text{H}}}$ into the coupled pressure diffusion partial differential equation,
\begin{align}
&\frac{\partial \Delta P(y,t)}{\partial t}-c_{hy}\frac{\partial^2 \Delta P(y,t)}{\partial y^2}=\Lambda\frac{\partial \Delta T(y,t)}{\partial t},\nonumber\\
&\Delta P(y,0)=0, \nonumber\\
&\Delta P(0,t)=\Delta P(\text{H},t)=0,
\end{align}
\noindent we obtain:
\begin{align}
\sum_{n=1}^\infty\frac{\partial \tilde{p}_n(t)}{\partial t}\sin{\lambda_n y}+c_{hy}\sum_{n=1}^\infty \lambda^2_n\tilde{p}_n(t)\sin{\lambda_n y}= \frac{2 \Lambda}{\text{H} \rho C}\sum_{n=1}^\infty\sin{\lambda_n y} \frac{\partial T_n{(t)}}{\partial t}&,\\
\intertext{where $T_n(t)$ is given as:}
T_n(t)=\int_0^t\int_{-\infty}^\infty g(y^\prime, t^\prime)\exp{\left[-\lambda_n^2 c_{th}(t-t^\prime)\right]}\sin{\lambda_n y^\prime}dy^\prime dt^\prime. \nonumber
\end{align}
\noindent Isolating each eigenfunction $\sin{\lambda_n y}$ we arrive at the following first order linear differential equations involving the unknown coefficient $\tilde{p}_n(t)$ and the loading coefficient $T_n(t)$ for each particular component of the solution series expansion.
\begin{align}
\frac{\partial \tilde{p}_n(t)}{\partial t}+c_{hy} \lambda^2_n\tilde{p}_n(t)=& \frac{2 \Lambda}{\text{H} \rho C}\frac{\partial T_n(t)}{\partial t},\;t\geq 0. 
\end{align}
\noindent Applying the Laplace transformation in the field of time:
\begin{align}
&s\tilde{P}_n(s)+c_{hy} \lambda^2_n\tilde{P}_n(s)=\frac{2 \Lambda}{\text{H} \rho C}\frac{s}{s+\lambda^2_n c_{th}}\int_{-\infty}^\infty G(y^\prime, s)\sin{\lambda_n y^\prime}dy^\prime .\\
&\tilde{P}_n(s)=\frac{2 \Lambda}{\text{H} \rho C}\frac{s}{(s+\lambda^2_n c_{th})(s+\lambda^2_n c_{hy})}\int_{-\infty}^\infty G(y^\prime, s)\sin{\lambda_n y^\prime}dy^\prime
\end{align}
\noindent Applying the inverse of the Laplace transform gives us:
\begin{align}
\tilde{p}_n(y,t) = \frac{2 \Lambda}{\text{H} \rho C }\int_0^t\int_{-\infty}^{\infty}g(y^\prime,t^\prime)\frac{c_{hy} \exp{\left[-\lambda^2_nc_{hy}(t-t^\prime)\right]}-c_{th} \exp{\left[-\lambda^2_nc_{th}(t-t^\prime)\right]}}{c_{hy}-c_{th}}\sin{\lambda_n y^\prime}dy^\prime dt^\prime
\end{align} 
\noindent Finally, in the series expansion $\Delta P(y,t)=\sum_{n=1}^\infty \tilde{p}_n(t)\sin{\lambda_n y}$ we move the summation under the integral sign and we obtain:
\begin{align}
\begin{aligned}
\Delta P(y,t) = \frac{2 \Lambda}{\text{H} \rho C }\int_0^t\int_{-\infty}^{\infty}&g(y^\prime,t^\prime)\\
&\sum_{n=1}^\infty\frac{c_{hy} \exp{\left[-\lambda^2_nc_{hy}(t-t^\prime)\right]}-c_{th} \exp{\left[-\lambda^2_nc_{th}(t-t^\prime)\right]}}{c_{hy}-c_{th}}\sin{\lambda_n y}\sin{\lambda_n y^\prime}dy^\prime dt^\prime
\end{aligned} \label{app E: eq: bounded Green's function Kernel press}.
\end{align}
\noindent We recognize the term in the second line of equation \eqref{app E: eq: bounded Green's function Kernel press} as the Green's function kernel of the coupled pressure diffusion partial differential equation. This expression has the added advantage that the influence of the thermal load on the pressure $\Delta P(y,t)=P(y,t)-P_0$ solution is straightforward. Noticing that for a general diffusion problem on a bounded domain under homogeneous Dirichlet boundary conditions the Green's function kernel is given by:
\begin{align}
G_{X11}(y,t;y^\prime,t^\prime,c) = \frac{2}{\text{H}}\sum_{n=1}^{\infty}\exp{\left[-\lambda^2_nc\frac{t-t^\prime}{\text{H}^2}\right]}\sin{\lambda_n y}\sin{\lambda_n y^\prime}.
\end{align}
The Green's function kernel of the coupled pressure differential equation on the bounded domain is then given as:
\begin{align}
G_{X11}(y,t;y^\prime,t^\prime,c_{th},c_{hy}) =\frac{ c_{hy}G_{X11}(y,t;y^\prime,t^\prime,c_{hy})-c_{th}G_{X11}(y,t;y^\prime,t^\prime,c_{th})}{c_{hy}-c_{th}}.
\end{align}
\noindent Finally, the pressure solution can be given as:
\begin{align}
\Delta P(y,t)=P(y,t)-P_0=\frac{\Lambda}{\rho C}\int_0^t\int_{-\infty}^\infty g(y^\prime,t^\prime)G_{X11}(y,t;y^\prime,t^\prime,c_{th},c_{hy})dy^\prime dt^\prime
\end{align}
This result agrees with the formula provided in \cite{Lee1987,Rice2006} for the unbounded domain.
\subsection{Moving thermal load, coupled pore fluid pressure Green's kernel for a unbounded domain.}
\noindent Here, we present the derivation of the Green's function kernel of the coupled pressure diffusion equation for an unbounded domain under moving thermal load. Note here, that the Green's function kernel is independent of the type of loading (stationary or moving), it depends on the kind of the differential operator and the boundary conditions. What differs here in the form of the Green's function kernel is the velocity dependence, since we want to connect the pressure evolution not with the stationary Green's function but with  the moving Dirac thermal load, that can be written as $g(y,t)=\frac{\dot{\delta}}{\rho C}\tau(t)\delta(y-vt)$. In essence we need to only prescribe the velocity dependence of $x^\prime=f(v,t^\prime)$ in the Green's function kernel for the unbounded domain under Dirichlet conditions $G_{X00}(y,t;y^\prime,t^\prime,c_{hy},c_{th})$. We provide a full description and then compare the results. The coupled system of temperature and pore fluid pressure diffusion equations in the unbounded domain is given by:
\begin{align}
&\frac{\partial \Delta T(y,t)}{\partial t}-c_{th}\frac{\partial^2 \Delta T(y,t)}{\partial y^2}=\frac{\dot{\delta}}{\rho C}\tau(t)\delta(y-vt),\;-\infty<y<\infty,\;0<t<\infty\nonumber\\
&\frac{\partial \Delta P(y,t)}{\partial t}-c_{hy}\frac{\partial^2 \Delta P(y,t)}{\partial y^2}=\Lambda\frac{\partial \Delta T(y,t)}{\partial t},\;-\infty<y<\infty,\;0<t<\infty,\nonumber\\
&\Delta T(y,0)=0,\nonumber\\
&\lim{\Delta T(y,t)}\|_{y=-\infty,y=\infty}=0,\nonumber\\
&\Delta P(y,0)=0,\nonumber\\
&\lim{\Delta P(y,t)}\|_{y=-\infty,x=\infty}=0 \label{app E: eq: system_of_pdes}
\end{align}
\noindent To account for the moving load we perform a change of variables on the original system \eqref{app E: eq: system_of_pdes}, setting $\xi=y-vt,\;\eta=t$ so that we attach a frame of reference to the moving load. In this case and by suitable application of the chain rule we can write:
\begin{align}
&\frac{\partial \Delta T(\xi,\eta)}{\partial \eta}-v\frac{\partial \Delta T}{\partial \xi}-c_{th}\frac{\partial^2 \Delta T(\xi,\eta)}{\partial \xi^2}=\frac{1}{\rho C}\tau(t)\delta(\xi),\;-\infty<\xi<\infty,\;0<\eta<\infty,\nonumber\\
&\frac{\partial \Delta P(\xi,\eta)}{\partial \eta}-v\frac{\partial \Delta P(\xi,\eta)}{\partial \xi}-c_{hy}\frac{\partial^2 \Delta P(\xi,\eta)}{\partial \xi^2}=\Lambda\frac{\partial \Delta T(\xi,\eta)}{\partial \eta},\;-\infty<\xi<\infty,\;0<\eta<\infty\nonumber\\
&\Delta T(\xi,0)=0,\nonumber
\end{align}
\begin{align}
&\lim{\Delta T(\xi,\eta)}\|_{\xi=-\infty,\xi=\infty}=0,\nonumber\\
&\Delta P(\xi,0)=0,\nonumber\\
&\lim{\Delta P(\xi,\eta)}\|_{\xi=-\infty,\xi=\infty}=0 \label{app E: eq: system_of_pdes_mov}
\end{align}
\noindent Applying a Fourier transform in space and a Laplace transform in time on the system of partial differential equations \eqref{app E: eq: system_of_pdes_mov} we obtain:
\begin{align}
&s \tilde{T}(k,s)-v(ik)\tilde{T}(k,s)-c_{th}(ik)^2\tilde{T}(k,s)=\frac{1}{\rho C}\tau(s),\nonumber\\
&s \tilde{P}(k,s)-v(ik)\tilde{P}(k,s)-c_{th}(ik)^2\tilde{P}(k,s)=\Lambda s \tilde{T}(k,s).\label{app E: eq: system_of_trans}
\end{align}
\noindent Solving the above algebraic system \eqref{app E: eq: system_of_trans} we obtain:
\begin{align}
&\tilde{T}(k,s) = \frac{1}{\rho C}\frac{\tau{(s)}}{s-v(ik)+c_{th}k^2},\\
&\tilde{P}(k,s) = \frac{\Lambda\tau(s)}{\rho C}\frac{s}{(s-v(ik)+c_{th}k^2)(s-v(ik)+c_{hy}k^2)}
\end{align}
\noindent Inverting the Laplace and then the Fourier transform yields:
\begin{align}
&\Delta T(y,t)= \frac{\dot{\delta}}{\rho C }\int_0^t\frac{\tau(t^\prime)}{2\sqrt{\pi c_{th}(t-t^\prime)}}\exp{\left[-\frac{(y-vt^\prime)^2}{4 c_{th}(t-t^\prime)}\right]} dt^\prime,\\
&\begin{aligned}
\Delta P(y,t)= \frac{\Lambda\dot{\delta}}{\rho C (c_{hy}-c_{th})}\int_0^t&\frac{\tau(t^\prime)}{2\sqrt{\pi(t-t^\prime)}}\left(\sqrt{c_{hy}}\exp{\left[-\frac{(y-vt^\prime)^2}{4 c_{hy}(t-t^\prime)}\right]} -\sqrt{c_{th}}\exp{\left[-\frac{(y-vt^\prime)^2}{4 c_{th}(t-t^\prime)}\right]} \right)\nonumber\\
dt^\prime.
\end{aligned}
\end{align}
\noindent  By inspection we note that these are the same expressions as the ones presented in \eqref{ch: 6 eq: G_X00_kernel}, where $y^\prime $ was replaced by $y^\prime=v t^\prime$ and $c=c_{th},\;\text{or}\;c=c_{hy}$ respectively.
\let\clearpage\relax

\section{Collocation Methodology \label{Appendix H}}
\subsection{Regular kernels}
\noindent In order to apply the collocation methodology to the linear Volterra integral equation of the second kind, \eqref{ch 6: eq: normalized_integral_equation}, we make use of the collocation methodology described in \cite{tang2008spectral}. The integral equation is given as:
\begin{align}
\bar{\tau}(\bar{t}) = 1-C\int_0^{\bar{t}}\bar{\tau}(\bar{t}^\prime)\bar{G}^\star(\bar{y},\bar{t};\bar{y}^\prime,\bar{t}^\prime)\|_{\bar{y}=\bar{y}^\prime} d\bar{t}^\prime,\; \bar{t}\in [0,\text{T}\frac{c_{th}}{\text{H}^2}] \label{app: H eq: integral equation1},
\end{align}
\noindent where $\bar{\text{T}}=\text{T}\frac{c_{th}}{\text{H}^2}$ is the final normalized time, and $\bar{\tau}(\bar{t})$ is the unknown function. We begin by performing a change of variables from $\bar{t}\in [0,\bar{\text{T}}]$ to $\bar{z}\in [-1,1]$. The change of variables reads:
\begin{align*}
\bar{t}=\bar{\text{T}}\frac{1+\bar{z}}{2},\;\bar{z}=\frac{2\bar{t}}{\bar{\text{T}}}-1
\end{align*}
\noindent The Volterra integral equation can then be written:
\begin{align}
U(\bar{z}) = 1-C\int_0^{\bar{\text{T}}\frac{1+\bar{z}}{2}}\bar{G}^\star(\bar{y},\bar{\text{T}}\frac{1+\bar{z}}{2};\bar{y}^\prime,\bar{t}^\prime)\|_{y=y^\prime}d\bar{t}^\prime,\; \bar{z}\in [-1,1] \label{app: H eq: integral equation2},
\end{align} 
\noindent where $U(\bar{z})=\tau(\bar{\text{T}}\frac{1+\bar{z}}{2})$. In order for the collocation method solution to converge exponentially we require that both the integral equation \eqref{app: H eq: integral equation2} and the integral inside \eqref{app: H eq: integral equation2} are expressed inside the same interval $[-1,1]$. To do this first we change the integration bounds from $\bar{t}^\prime\in[0,\bar{\text{T}}\frac{1+\bar{z}}{2}]$ to $s\in [-1,\bar{z}]$.
\begin{align}
U(\bar{z}) = 1-C\int_0^{\bar{z}}K(\bar{y},\bar{z};\bar{y}^\prime,s)\|_{\bar{y}=\bar{y}^\prime}U(s)ds,\; \bar{z}\in [-1,1], \label{app: H eq: integral equation3},
\end{align} 
\noindent where $K(\bar{y},\bar{z};\bar{y^\prime},s)=\frac{\bar{\text{T}}}{2}\bar{G}^\star(\bar{y},\frac{\bar{\text{T}}}{2}(\bar{z}+1);\bar{y^\prime},\frac{\bar{\text{T}}}{2}(s+1))$. Next, we set the $N+1$ collocation points $\bar{z}_i\in [-1,1]$ and corresponding weights $\omega_i$ according to the Clenshaw-Curtis quadrature formula. The integral equation \eqref{app: H eq: integral equation3} must hold at each $\bar{z}_i$:
\begin{align}
U(\bar{z}_i) = 1-C\int_0^{\bar{z}_i}K(\bar{y},\bar{z}_i;\bar{y}^\prime,s)\|_{\bar{y}=\bar{y}^\prime}U(s)ds,\; i\in [0,N], \label{app: H eq: integral equation4},
\end{align}
\noindent The main hindrance in solving equation \eqref{app: H eq: integral equation4} accurately, is the calculation of the integral with variable integration bounds. For small values of $\bar{z}_i$, the quadrature provides little information for $U(s)$. We handle this difficulty by yet another variable change where we transfer the integration variable $s\in [-1,\bar{z}_i]$ to $\theta\in [-1,1]$ via the transformation:
\begin{align}
s(\bar{z},\theta)=\frac{1+\bar{z}}{2}\theta+\frac{\bar{z}-1}{2},\;\theta \in [-1,1].
\end{align}
Thus, equation \eqref{app: H eq: integral equation4} is transformed into:
\begin{align}
U_i+C\frac{1+\bar{z}_i}{2}\sum\limits_{j=0}^N{}^\prime u_j\sum\limits_{p=0}^N K(\bar{y},\bar{z}_i,\bar{y}^\prime,s(\bar{z}_i,\theta))\|_{\bar{y}=\bar{y}^\prime}U(s(\bar{z}_i,\theta_p))\omega_j=1,\; i\in [0,N]
\end{align}
\noindent In order to apply the collocation method according to the Clenshaw-Curtis quadrature, we express the solution $U(s(\bar{z}_i,\theta_p))$ with the help of Lagrange interpolation polynomials $P_j(s(\bar{z}_i,\theta_p))$ as a series: $U(s(\bar{z}_i,\theta_p)) \sim \sum  \limits_{k=0}^N{}^\prime U_jP_j(s(\bar{z}_i,\theta_j))$,
\begin{align}
U_i+\frac{1+\bar{z}_i}{2}\sum\limits_{j=0}^N{}^\prime U_j\left(\sum\limits_{p=0}^N K(y,\bar{z}_i;\bar{y}^\prime,s(\bar{z}_i,\theta))\|_{\bar{y}=\bar{y}^\prime}P_j(s(\bar{z}_i,\theta_p))\omega_p\right)=1,\; i\in [0,N]
\end{align}
\noindent where, $P_j(s(\bar{z_i},\theta_p)),\;\sum\limits_{j=0}^N{}^\prime$ have been defined in the main text (see equations \eqref{ch: 6 eq: barycentric formula} and \eqref{ch: 6 eq: mod sum}). In order to assure an exponential degree of convergence, we choose the set of Gauss Chebyshev quadrature points for the numerical evaluation of the integral $\{\theta_j\}_{j=0}^N$, to coincide with the set of collocation points $\{\bar{z}_j\}_{j=0}^N$, where the integral equation is evaluated. Rearranging the terms and applying Einstein's summation over repeated indices yields the system of algebraic equations:
\begin{align}
(\delta_{ij}+A_{ij})U_j=g(\bar{z}_i),
\end{align}
\noindent where, $A_{ij} =\frac{1+\bar{z}_i}{2}\sum\limits_{j=0}^N{}^\prime\left(\sum\limits_{p=0}^N K(\bar{y},\bar{z}_i;\bar{y}^\prime,s(\bar{z}_i,\theta))P_j(s(\bar{z}_i,\theta_p))\omega_p\right),\; g(\bar{z}_i)=1$ and $U_j$ the unknown quantities. Because Lagrange interpolation was assumed, the interpolation coefficients $U_j$ calculated at each $\bar{x}_j$ are also the value of the interpolation at $\bar{z}_j$.
\subsection{Singular kernels}
\noindent When the kernel of the integral equation \eqref{ch 6: eq: normalized_integral_equation} involves a singularity (see equation \eqref{ch 6: eq: normalized_unbounded_kernel}), we cannot use the Clenshaw-Curtis quadrature rule in its original form because the quadrature requires the values of the function at a position, where the kernel evaluates to infinity. For this reason a different quadrature strategy needs to be implemented. Here, based on the work of \cite{tang2008spectral} we apply the Gauss-Chebyshev quadrature rule. This quadrature rule involves the values of the function at the zeros of the $N$-th degree Chebyshev polynomial of the first kind $\{\bar{z}^\prime_i\}$. The quadrature can then be successfully calculated, because the new set of integration points, $\{\bar{z}^\prime_i\}$, does not involve the ends of the interval [-1,1]. However, since the Chebyshev polynomials of the first kind were used, we need to take into account the specific weight function $w(\bar{z})=\sqrt{1-\bar{z}^2}$ under which the Chebyshev polynomials of the first kind are orthogonal on the interval [-1,1], namely:
\begin{align} \label{orthogonality_condition}
\int_{-1}^1\frac{T_{i}(\bar{z})T_{j}(\bar{z})}{\sqrt{1-\bar{z}^2}}d\bar{z}=\begin{cases}&1,\;i=j\\ 
&0,\;i\neq j\end{cases},
\end{align}
\noindent moreover, due to the change in the evaluation set, $\{\bar{z}^\prime_i\}$, the formula for the calculation of the Lagrange interpolation is given by:
\begin{align} 
U(s(\bar{z},\theta_p))=\sum\limits_{j=0}^NU(\bar{z}^\prime_j)F_j(s(\bar{z},\theta_p))
\end{align}
\noindent where $F_j(s(\bar{z},\theta_p)$ are the Lagrange cardinal polynomials. We note that the formula for the Lagrange cardinal polynomials changes due to the change of the interpolation nodes $\{\bar{z}^\prime_j\}^N_{j=0}$. Taking advantage of the orthogonality condition the cardinal polynomials $F_j(s(\bar{z},\theta_p))$ are given by:
\begin{align}
F_j(s(\bar{z}^\prime_i,\theta_p))=\sum_0^N\alpha_{p,j}T_{p}(s(\bar{z}^\prime_i,\theta_p)),
\end{align}   
\noindent where, again due to orthogonality, $\alpha_{p,j}$ is given by:
\begin{align}
\alpha_{p.j}=T_{p}(\bar{z}^\prime_j)\omega_j/\gamma_p,\\
\gamma_p=\begin{cases}&\pi,\;j=0\\
					  &\frac{\pi}{2},\;j\neq 0 
					  \end{cases}
\end{align}
\noindent The final discretized form of the integral equation \eqref{ch 6: eq: normalized_integral_equation} is then given by:
\begin{align}
U_i+\frac{1+\bar{z}_i}{2}\sum\limits_{j=0}^NU_j\left(\sum\limits_{p=0}^N K(y,\bar{z}_i;\bar{y}^\prime,s(\bar{z}_i,\theta))\|_{\bar{y}=\bar{y}^\prime}F_j(s(\bar{z}_i,\theta_p))\sqrt{1-\theta^2_p}\omega_p\right)=1,\; i\in [0,N].
\end{align}
\noindent we note here that the term $\sqrt{1-\theta^2_p}$, accounts for the weight function present in the orthogonality condition.

\nocite{*} 
\typeout{}

\bibliography{/home/alexandrosstathas/Documents/PhD_topics/PAPER_III/biblio/bibliography_clean.bib}

\end{document}